\documentclass[a4paper,11pt]{article}

\usepackage{jinstpub} 
\usepackage{color} % colors
\usepackage{xcolor} % colors
\usepackage{graphicx} % graphics
\usepackage{stmaryrd}
\usepackage{tikz} % custom drawing
\usetikzlibrary{shapes,arrows,positioning,automata,backgrounds,calc,er,patterns,decorations.markings, shapes.misc} % tikz instances
\usepackage[compat=1.1.0]{tikz-feynman} % feynman diagrams
\usepackage{circuitikz} % draw circuits with latex :-)
\ctikzset{bipoles/length=.8cm}
\usepackage{subcaption} % contains the 'subfigure' environment
\usepackage{physics} % pyhsics symbols
\usepackage{units} % proper units in formulae ( usage: \unit[<value>]{<dimension>} or \unitfrac[<value>]{<numerator>}{<denominator>} )
\usepackage{mathtools} % more symbols
\usepackage{cleveref}
\usepackage{wrapfig} % make inline figures wrapped by text
\usepackage{tabularx}
\usepackage{booktabs}
\usepackage{makecell}
\usepackage{multirow}
\usepackage{multicol}
\usepackage[style=chem-angew,
            maxbibnames=1,
            url=false,
            doi=false,
            isbn=false,
            ]{biblatex}
\AtEveryBibitem{%
  \clearfield{note}%
  \clearfield{doi}%
  \clearlist{language}%
}
\Crefname{tabular}{tab.}{tabs.}
\Crefname{table}{tab.}{tabs.}

\title{Optical calibration systems of the Pacific Ocean Neutrino Experiment}

\author[a]{M.~Agostini,}
\author[a]{A.~Alexander Wight,}
\author[b]{M.~Altomare,}
\author[c]{K.~Ba\c{s},}
\author[d]{N.~Baily,}
\author[e,f]{P.~S.~Barbeau,}
\author[d]{A.~J.~Baron,}
\author[c]{S.~Bash,}
\author[c]{C.~Bellenghi,}
\author[c]{M.~Boehmer,}
\author[c]{M.~Brandenburg,}
\author[d]{P.~Bunton,}
\author[e,f]{N.~Cedarblade-Jones,}
\author[a]{B.~Crudele,}
\author[b]{M.~Danninger,}
\author[g]{T.~DeYoung,}
\author[b]{A.~G\"{a}rtner,}
\author[g]{J.~Garriz,}
\author[b]{D.~Ghuman,}
\author[c]{L.~Ginzkey,}
\author[b]{T.~Glukler,}
\author[c]{V.~Gousy-Leblanc,}
\author[b]{D.~Grant,}
\author[b]{A.~Grimes,}
\author[h]{C.~Haack,}
\author[b]{R.~Hall,}
\author[i]{R.~Halliday,}
\author[d]{D.~Hembroff,}
\author[b,h,*]{F.~Henningsen,}
\author[d]{M.~Herle,}
\author[h]{O.~Janik,}
\author[b]{H.~Johnson,}
\author[j]{W.~Kang,}
\author[k]{S.~Karanth,}
\author[c]{T.~Kerscher,}
\author[d]{S.~Kerschtien,}
\author[k]{K.~Kopa\'{n}ski,}
\author[h]{C.~Kopper,}
\author[b]{P.~Krause,}
\author[l]{C.~B.~Krauss,}
\author[j]{N.~Kurahashi,}
\author[h]{C.~Lagunas Gualda,}
\author[d]{A.~Lam,}
\author[d]{T.~Lavallee,}
\author[c]{K.~Leism\"{u}ller,}
\author[c]{R.~Li,}
\author[c]{S.~Loipolder,}
\author[d]{C.~Magee,}
\author[c]{S.~Magel,}
\author[k]{P.~Malecki,}
\author[l]{T.~Martin,}
\author[l]{A.~Maunder,}
\author[b,m]{C.~Miller,}
\author[l]{N.~Molberg,}
\author[l]{R.~Moore,}
\author[c]{B.~N\"{u}hrenb\"{o}rger,}
\author[b]{B.~Nichol,}
\author[k]{W.~Noga,}
\author[c]{R.~\O{}rs\o{}e,}
\author[c]{L.~Papp,}
\author[g]{V.~Parrish,}
\author[c]{P.~Pfahler,}
\author[c]{J.~Pflanz,}
\author[d]{B.~Pirenne,}
\author[d]{E.~Price,}
\author[n,o]{A.~Rahlin,}
\author[l]{M.~Rangen,}
\author[c]{E.~Resconi,}
\author[l]{S.~Robertson,}
\author[g]{M.~F.~Rodriguez-Pilco,}
\author[g]{D.~Salazar-Gallegos,}
\author[c]{A.~Scholz,}
\author[h]{L.~Schumacher,}
\author[k]{S.~Sharma,}
\author[p]{B.~R.~Smithers,}
\author[c]{C.~Spannfellner,}
\author[b,*]{J.~Stacho,}
\author[q]{I.~Taboada,}
\author[c]{K.~Tchiorniy,}
\author[g]{J.~P.~Twagirayezu,}
\author[g]{M.~Un Nisa,}
\author[l]{B.~Veenstra,}
\author[q]{M.~Velazquez,}
\author[c]{L.~von der Werth,}
\author[g]{C.~Weaver,}
\author[g]{N.~Whitehorn,}
\author[m]{B.~Winnicky-Lewis,}
\author[c]{L.~Winter,}
\author[k]{R.~Wro\'{n}ski,}
\author[l]{J.~P.~Ya\~{n}ez,}
\author[j]{S.~Yun-C\'{a}rcamo,}
\author[e,f]{A.~Zaalishvili}

\affiliation[a]{Department of Physics and Astronomy, University College London, Gower Street, London, WC1E 6BT, UK}
\affiliation[b]{Department of Physics, Simon Fraser University, 8888 University Drive, Burnaby, BC, Canada, V5A 1S6}
\affiliation[c]{Physik-Department, Technische Universit\"{a}t M\"{u}nchen, D-85748 Garching, Germany}
\affiliation[d]{Ocean Networks Canada, University of Victoria, Victoria, BC, Canada}
\affiliation[e]{Department of Physics, Duke University, Durham, NC, 27708, USA}
\affiliation[f]{Triangle Universities Nuclear Laboratory, Durham, NC, 27708, USA}
\affiliation[g]{Department of Physics and Astronomy, Michigan State University, East Lansing, MI, 48824, USA}
\affiliation[h]{Erlangen Centre for Astroparticle Physics, Friedrich-Alexander-Universit{\"a}t Erlangen-N\"{u}rnberg, D-91058 Erlangen, Germany}
\affiliation[i]{Department of Physics, Elmhurst University, 190 S. Propsect Ave, Elmhurst, IL, 60126, USA}
\affiliation[j]{Department of Physics, Drexel University, 3141 Chestnut Street, Philadelphia, PA, 19104, USA}
\affiliation[k]{Institute of Nuclear Physics, Polish Academy of Sciences, Krakow, Poland}
\affiliation[l]{Department of Physics, University of Alberta, Edmonton, AB, Canada, T6G 2E1}
\affiliation[m]{Department of Physics and Astronomy, University of Victoria, 3800 Finnerty Road, Victoria, BC, Canada, V8P 5C2}
\affiliation[n]{Department of Astronomy and Astrophysics, University of Chicago, 5640 South Ellis Avenue, Chicago, IL, 60637, USA}
\affiliation[o]{Kavli Institute for Cosmological Physics, University of Chicago, 5640 South Ellis Avenue, Chicago, IL, 60637, USA}
\affiliation[p]{TRIUMF, 4004 Wesbrook Mall, Vancouver, BC, Canada, V6T 2A3}
\affiliation[q]{School of Physics and Center for Relativistic Astrophysics, Georgia Institute of Technology, Atlanta, GA, 30332, USA}

\emailAdd{felix.henningsen@fau.de}
\emailAdd{jakub\_stacho@sfu.ca}

\abstract{
This work presents the design and performance characterization of the optical calibration systems produced for the Pacific Ocean Neutrino Experiment (P-ONE), which target gain, energy and time calibration in the detector. These systems include novel light-pulse driver circuitry based on gallium nitride field-effect transistor technology and its application to directional and isotropic, self-monitoring optical calibration instruments. A total of 330 directional light pulsers and two isotropic, 17-inch calibration modules (P-CALs) were produced for the first P-ONE line. We present the designs and performance of both the directional and isotropic calibration devices and perform detailed optical characterizations of both full-production batches. In a wavelength range of \unit[$365 - 520$]{nm}, our developed driver circuits achieve emission intensities up to \unit[$10^{11}$]{photons} and pulse widths as small as \unit[1.4]{ns}, respectively. Light-pulse drivers and self-monitoring electronics in the P-CAL were characterized using the same experimental setup, and the instrument's optical-isotropy design was optimized in combination with a dedicated GEANT4-based simulation framework. The optimized P-CAL achieves a simulated isotropy grade of \unit[$1.00 \pm 0.01$]{} across the entire \unit[$4\pi$]{solid angle} range. These simulation investigations were explicitly confirmed by dedicated measurements in both air and water using two independent experimental setups, and we report the results. With this, a detailed performance estimate for deployed P-CAL modules in P-ONE was possible.
}


\definecolor{rev1}{rgb}{0,0,0}
\definecolor{rev2}{rgb}{0,1,0}
\definecolor{rev3}{rgb}{0,0,1}

\collaboration{\includegraphics[height=17mm]{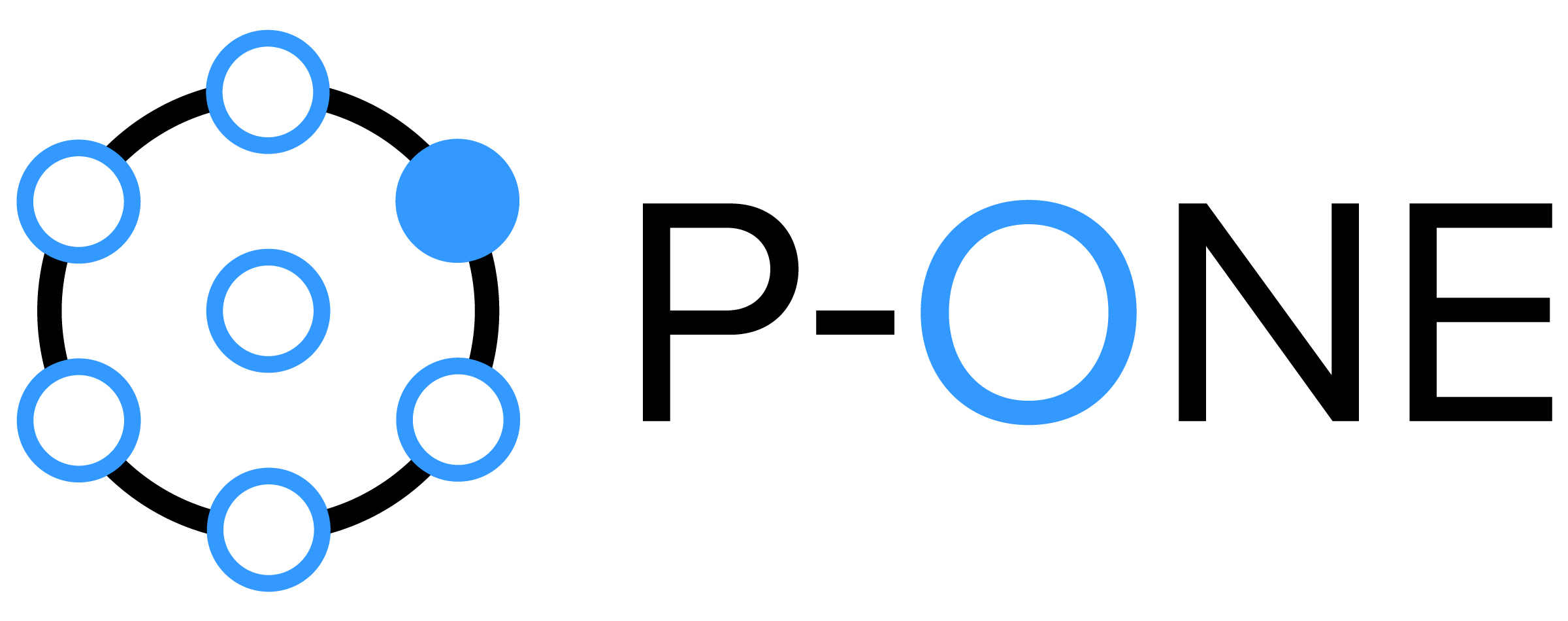}\\[6pt]
  The P-ONE collaboration}
\begin{document}
\maketitle
\flushbottom

%%%% INTRO %%%%%%%%%%%%%
\section{Introduction}
\label{sec:intro}
The Pacific Ocean Neutrino Experiment (P-ONE) is a new large-volume neutrino detector proposed for the deepsea of the Pacific Ocean~\cite{agostini_pacific_2020}. In collaboration with the existing sub-sea infrastructure operated and maintained by Ocean Networks Canada (ONC), P-ONE aims to instrument more than a cubic-kilometer of ocean water with photosensitive instrumentation to detect Cherenkov light signatures produced in high-energy neutrino interactions. Detecting these in time and space with sensor arrays allows identifying their energy and direction. The latter eventually allows pointing the detected neutrinos back to their sources, ultimately enabling neutrino astronomy.

The ocean is a dynamic environment that makes the installation and operation of large-scale detectors extremely complex. P-ONE is a novel initiative tackling this challenge. P-ONE's main objective is to combine leading-edge technologies with ONC's proven deep-sea infrastructure and expertise to establish a leading high-energy neutrino detection facility in the Pacific Ocean. The P-ONE array will be located about \unit[300]{km} West of Vancouver Island on the flat abyssal plain of the Cascada Basin in a water depth of about \unit[2.7]{km}. The array will comprise multiple 1 km-tall vertical detection lines, kept upright with subsea floats and connected to ONC's seafloor infrastructure. The first line, referred to as P-ONE-1, will carry 20 instruments, comprising 18 regular optical modules (P-OMs) and two calibration modules (P-CALs). The latter sit approximately one and two thirds along the line height for best optical coverage. P-OMs are equipped with 16 Hamamatsu R14374-10 photomultiplier tubes (PMTs), distributed between two hemispheres; P-CALs replace four PMTs in each hemisphere with a diffuse flasher system~\cite{spannfellner_design_2023}. A mounting frame in each hemisphere holds the PMTs in place and allows for the integration of calibration devices. Leveraging multi-PMT modules with full-waveform digitization, sub-nanosecond time synchronization, extensive calibration systems, and existing ONC infrastructure, P-ONE aims to detect astrophysical neutrinos in the TeV to PeV energy range; see~\cite{park_canadian_2025} for a detailed review of the P-ONE science objectives.

Two pathfinder experiments, STRAW~\cite{boehmer_straw_2019} and STRAW-b~\cite{holzapfel_straw-b_2023}, were deployed in 2018 and 2020, respectively, to characterize the optical properties of the P-ONE site. With its \unit[150]{m} length, STRAW measured an attenuation length of approximately \unit[30]{m} at \unit[450]{nm}~\cite{bailly_two-year_2021} in its accessible water column, demonstrating sufficient transparency between \unit[$350-500$]{nm} for a large-volume Cherenkov detector. Together with STRAW, the \unit[450]{m}-long STRAW-b further contributed to the characterization of optical background light, including Potassium-40 (K-40)~\cite{bailly_two-year_2021} and bioluminescence~\cite{holzapfel_pathfinders_2023}. 

\begin{figure}[h!]
    \centering
    % \vspace{-12pt}
    \includegraphics[width=.6\textwidth]{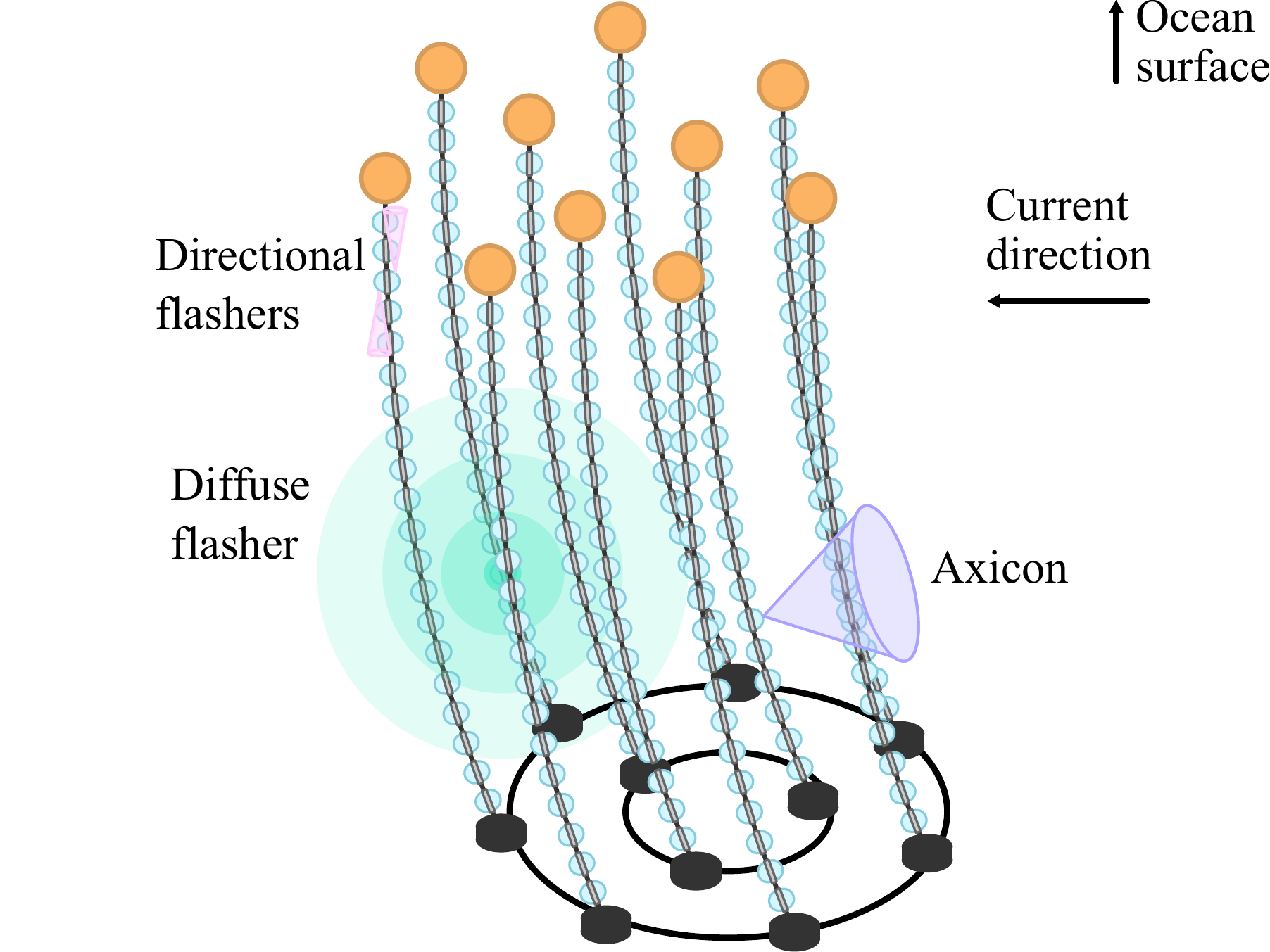}
    \caption{Conceptual view of the optical calibration systems in a cluster of P-ONE detection lines together with illustrations of the emission profiles of directional, isotropic, and axicon flashers.}
    \label{fig:pone-systems-summary}
    \vspace{-12pt}
\end{figure}

Continuous detector calibration and monitoring are essential to maintain precision throughout P-ONE's lifetime. Therefore, P-ONE will be instrumented with a robust suite of calibration systems, including an acoustic positioning system~\cite{agostini_prototype_2025} and optical light sources. The latter are the primary focus of this article and come in directional (\cref{sec:directional}) and isotropic (\cref{sec:pcal}) variants: both responsible for inter-module measurements of water absorption, scattering, and dispersion, as well as time synchronization, biofouling, and sedimentation. Biofouling and sedimentation refer to the accumulation of particulate matter and organisms on the glass surface~\cite{aghaei_long_2025}. Timing is critical for the angular resolution of neutrino reconstruction so we target a precision better than the PMT transit-time spread of around \unit[1]{ns}. The first P-ONE line is instrumented with two P-CALs and roughly 300 directional flashers. P-CALs will additionally be used for geometry calibration once more lines are deployed. Similar optical systems have been implemented in other experiments including IceCube~\cite{aartsen_measurement_2013,henningsen_self-monitoring_2020,abbasi_-situ_2022}, KM3NeT~\cite{aiello_nanobeacon_2022}, and Baikal-GVD~\cite{avrorin_baikal-gvd_2021}. A summary for P-ONE is depicted in \cref{fig:pone-systems-summary} and shows both flasher types, and a special variant of the directional flasher with an axicon lens.
\par%\noindent
%

%%%% Flashers %%%%%%%%%%%%
\section{Directional flasher}
\label{sec:directional}
The directional light pulsers (or flashers) developed for the first P-ONE detection line are intended for water and timing calibration. Because the entire kilometer of water column needs to be calibrated and monitored, flashers are included in all modules. This section outlines the optical pulser design and its performance characteristics.

%%%%%%%%%%%%%%%%%%%%%%%%%%%%%%%%%%%%
% REQUIREMENTS
%%%%%%%%%%%%%%%%%%%%%%%%%%%%%%%%%%%%
\subsection{Requirements}
The goals for the directional flashers are to perform timing synchronization and measure scattering, absorption, and dispersion properties of the water. Secondary calibrations include module efficiency and PMT gain. This implies that flashers should ideally illuminate multiple modules along the line to improve statistics and constrain systematic uncertainties.

The system distributes a total of 16 flashers in three orientations within each optical module: 6x up, 6x down, and 4x perpendicular to the line. Here, "up" refers to flashers pointing toward the ocean surface and "down" toward the seafloor. Flasher wavelengths were selected in the window of water transparency between \unit[$350-520$]{nm}, known from STRAW measurements~\cite{bailly_two-year_2021}. With these orientations and spectra, each pair of modules forms a combination of up- and down-facing flashers, allowing cross-calibration of the same intermediate water volume and optical modules, while providing maximum redundancy. In addition, the intensity ratio measured between such pairs allows assessing biofouling and sedimentation on the glass pressure housing over time, which is expected to be more prominent on the surface-facing side of the instruments. Flashers pointing perpendicular to the line are initially used for back-scatter measurements, and later for inter-line time synchronization. P-CALs are only equipped with four directional flashers, two pointing up and two pointing down, operating at central wavelengths of \unit[365]{nm}, \unit[450]{nm} (up and down), and \unit[520]{nm}. The remaining flasher channels in the P-CAL are occupied by the diffuse system (see \cref{sec:pcal}).

\begin{figure}[h!]
    \centering
    \begin{subfigure}{.46\textwidth}
      \centering
      \includegraphics[width=.95\textwidth]{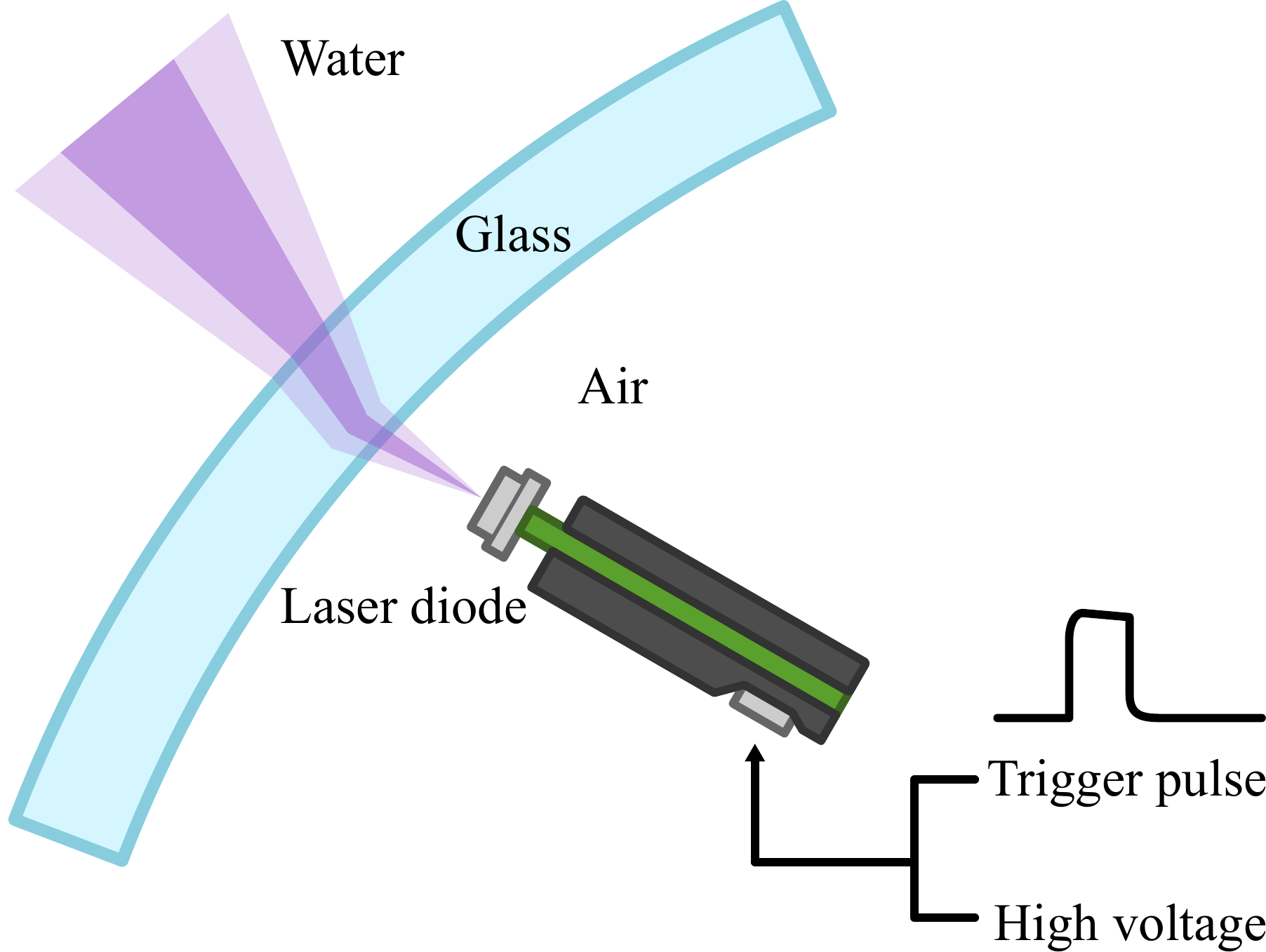}
      \caption{Directional flasher concept.}
      % \label{fig:setup-lab-a}
    \end{subfigure}%
    \hspace{1mm}
    \begin{subfigure}{.46\textwidth}
      \centering
      \includegraphics[width=.95\textwidth]{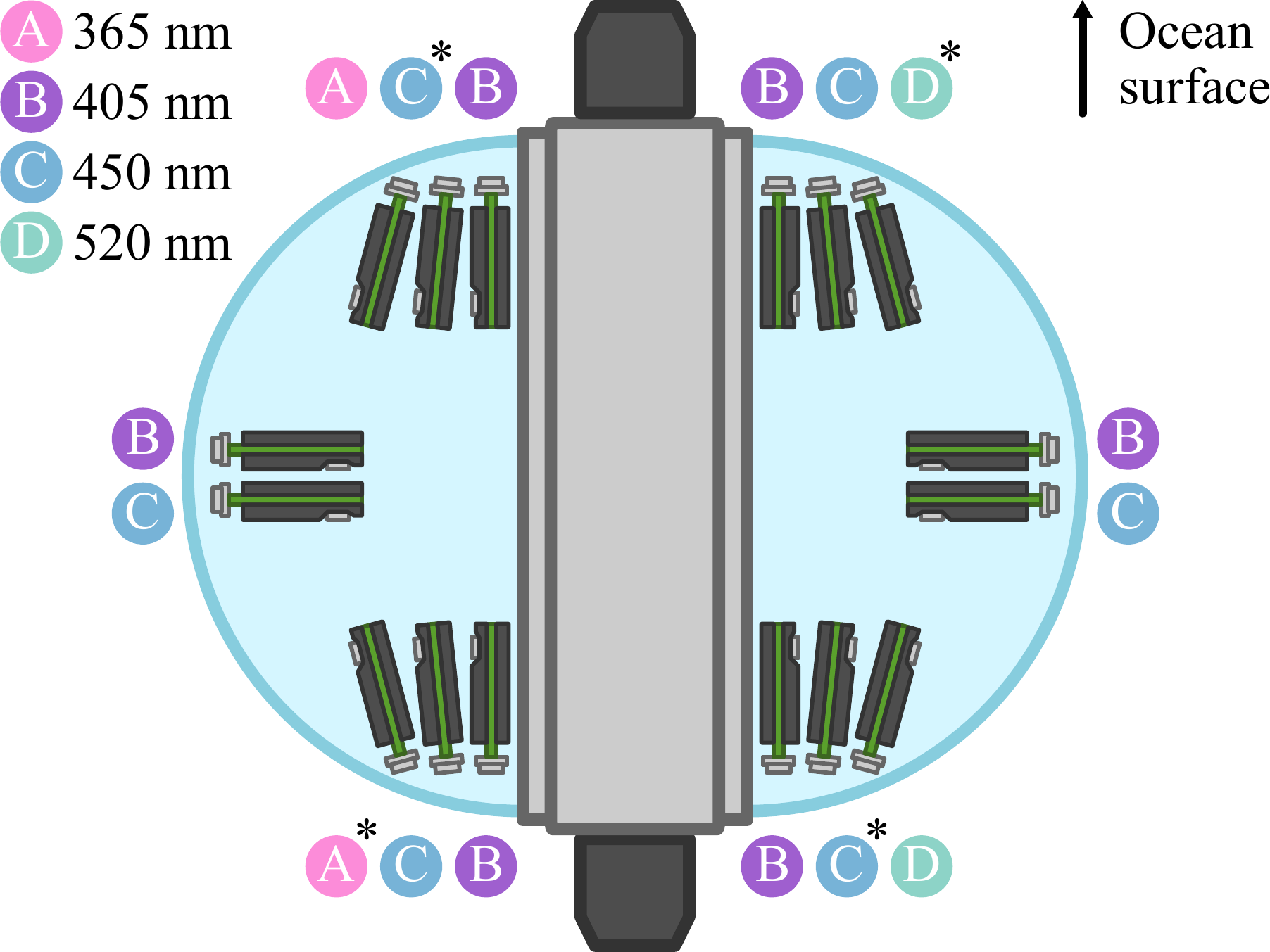}
      \caption{Optical module flasher layout.}
      % \label{fig:setup-lab-b}
    \end{subfigure}
    \caption{Concept of directional flashers in P-ONE-1. The images show \textbf{(a)} details of the directional flasher concept and \textbf{(b)} their integrated layout in P-OM and P-CAL (*); adapted from~\cite{ghuman_situ_2025}.}
    \label{fig:sketch-flashers}
\end{figure}

A sketch of this layout and wavelength selection for the P-OM is shown in \cref{fig:sketch-flashers}. With 18 P-OMs and two P-CALs, this results in a total of 296 directional flashers distributed along the full kilometer of the first P-ONE mooring line. 
\par%\noindent
The properties of the optical pulse critically determine calibration precision. For example, knowledge of the emitted light intensity translates into calibration precision for water absorption and attenuation, and efficiency measurements; timing properties of the pulse affect measurement precision for time synchronization, dispersion, and scattering. STRAW results indicate that optical transmission at the P-ONE site is highest in the spectral window between \unit[$350-520$]{nm}~\cite{bailly_two-year_2021}, making center wavelengths in that window preferable. A minimum light yield requirement of \unit[$10^{8}-10^{9}$]{photons per pulse} ensures sufficient photon statistics up to a distance of at least two modules (or \unit[100]{m}). The pulse full-width at half maximum (FWHM) requirement of about \unit[3]{ns} allows measuring the group velocity (and, with it, wavelength-dependent dispersion) in water with a precision of approximately \unit[1]{\%} at a distance of \unit[50]{m}. Since the latter drives photon timing, it directly impacts the neutrino angular reconstruction precision and is therefore a critical measurement.

%%%%%%%%%%%%%%%%%%%%%%%%%%%%%%%%%%%%
% DESIGN
%%%%%%%%%%%%%%%%%%%%%%%%%%%%%%%%%%%%
\subsection{Light pulser design}
The light pulser design closely follows the sub-ns configuration in~\cite{henningsen_picosecond_2023} with the EPC8004 Gallium Nitride field-effect transistor (GaNFET) in combination with the Texas Instruments LMG1020 gate driver. This design enables stable, high-current, sub-nanosecond pulsing with selectable input pulse widths and a long driver lifetime. Pulsing is possible with both light-emitting diodes (LEDs) and laser diodes (LDs) at high frequencies.

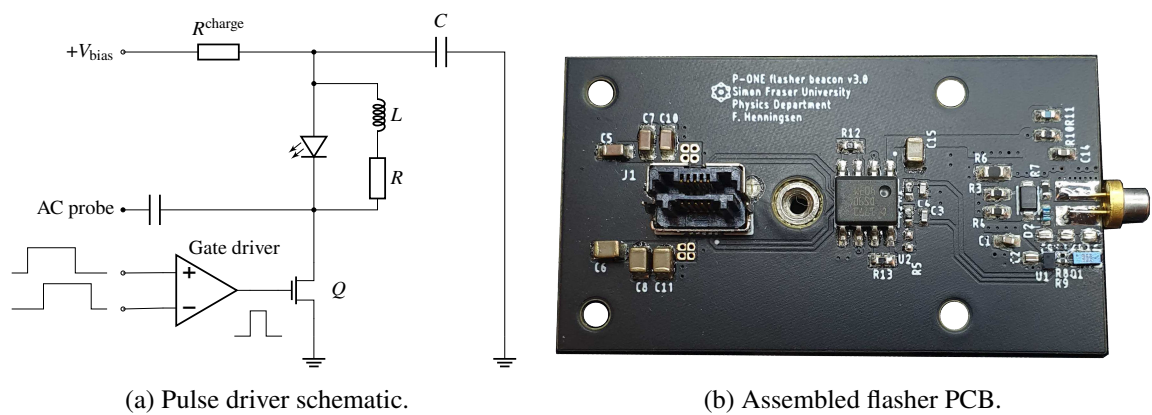
\begin{figure}[h!]
    \centering
    \vspace*{1mm}
    \begin{subfigure}{.45\textwidth}
      \centering
      \resizebox{0.99\textwidth}{!}{\begin{circuitikz}[american voltages]
    \ctikzset{resistor = european}
        % origin
		\draw
            (0,0) node[ocirc] {}
            (0,-0.57) node[ocirc] {};
		% VCC
  	\draw
            (0,3.5) node[ocirc] {}
            (0,3.5) node[left] {\footnotesize ~$+V_{\text{bias}}$};
		% trigger input part
  	\draw
            (-1.75,-0.57)
            to [short] (-1.25, -0.57)
            to [short] (-1.25,-0.17)
            to [short] (-0.5,-0.17)
            to [short] (-0.5,-0.57)
            to [short] (-0.25,-0.57);
	    \draw
            (-1.75,0)
            to [short] (-1.5, 0)
            to [short] (-1.5,0.4)
            to [short] (-0.75,0.4)
            to [short] (-0.75,0)
            to [short] (-0.25,0);
  	% trigger output
  	\draw
            (1.75,-1)
            to [short] (2,-1)
            to [short] (2,-0.6)
            to [short] (2.25,-0.6)
            to [short] (2.25,-1)
            to [short] (2.5,-1);
  	% LMG
  	\draw 
            (1.3125,-0.275) node[op amp,scale=1,yscale=-1] (lmg) {}
            (1.75, 0.4) node[] {\footnotesize Gate driver};
  	% gan fet
  	\draw 
            (3,-0.275) node[nmos,scale=1](npn){}
  		(3.4, -0.275) node[] {\footnotesize $Q$};
  	% ground
  	\draw
            (3,-1) node[ground,scale=1.5] (gnd) {}
            (6,-1) node[ground,scale=1.5] (gnd2) {};
  	% circuit paths
  	\draw
            % lmg
            (0,0) 
            to [short] (lmg.+)
            (0,-0.57) 
            to [short] (lmg.-)
            (lmg.out) to (npn.G)
            % fet / led
            (npn.D)
            to (3,1)
            to [led, invert] (3,3)
            to (3,3.5)
            to [resistor, l_={\footnotesize $R^\text{charge}$}] (0,3.5)
            % RL tail cutter
            (3,3)
            to (4,3)
            to [inductor,l^={\footnotesize$L$}] (4,2)
            to [resistor,l^={\footnotesize$R$}] (4,1)
            to (3,1)
            % fet / gnd
            (npn.S) to [short] (gnd)
            % C / gnd
            (3,3.5) to (4,3.5)
            (4,3.5)
            to [capacitor, a^={\footnotesize$C$}] (6,3.5)
            to (gnd2)
            % vdrain monitoring
            (3,1)
            to (1,1)
            to [capacitor] (0,1) node[circ] {}
            (0,1) node[left] {{\footnotesize AC probe}};
  	\draw
            (3,3.5) node[circ] {}
            (3,3) node[circ] {}
            (3,1) node[circ] {};
        % charging current
        %\draw [-stealth,red,thick](2.5,3.5) -- (3.5,3.5);
        %\draw (3, 4) node[red] {\footnotesize $I_{\text{charge}}$};
        % discharging current
        %\draw [-stealth,red,thick](2.25, 2) -- (2.25,1);
        %\draw (1.5, 1.5) node[red] {\footnotesize $I_{\text{discharge}}$};
        % grid
        %\draw[step=1cm,gray,very thin,opacity=0.2] (-2,-2) grid (6,4);
\end{circuitikz}}
      \caption{Pulse driver schematic.}
      % \label{fig:setup-lab-a}
    \end{subfigure}%
    \hspace{2mm}
    \begin{subfigure}{.53\textwidth}
      \centering
      \includegraphics[width=.99\textwidth]{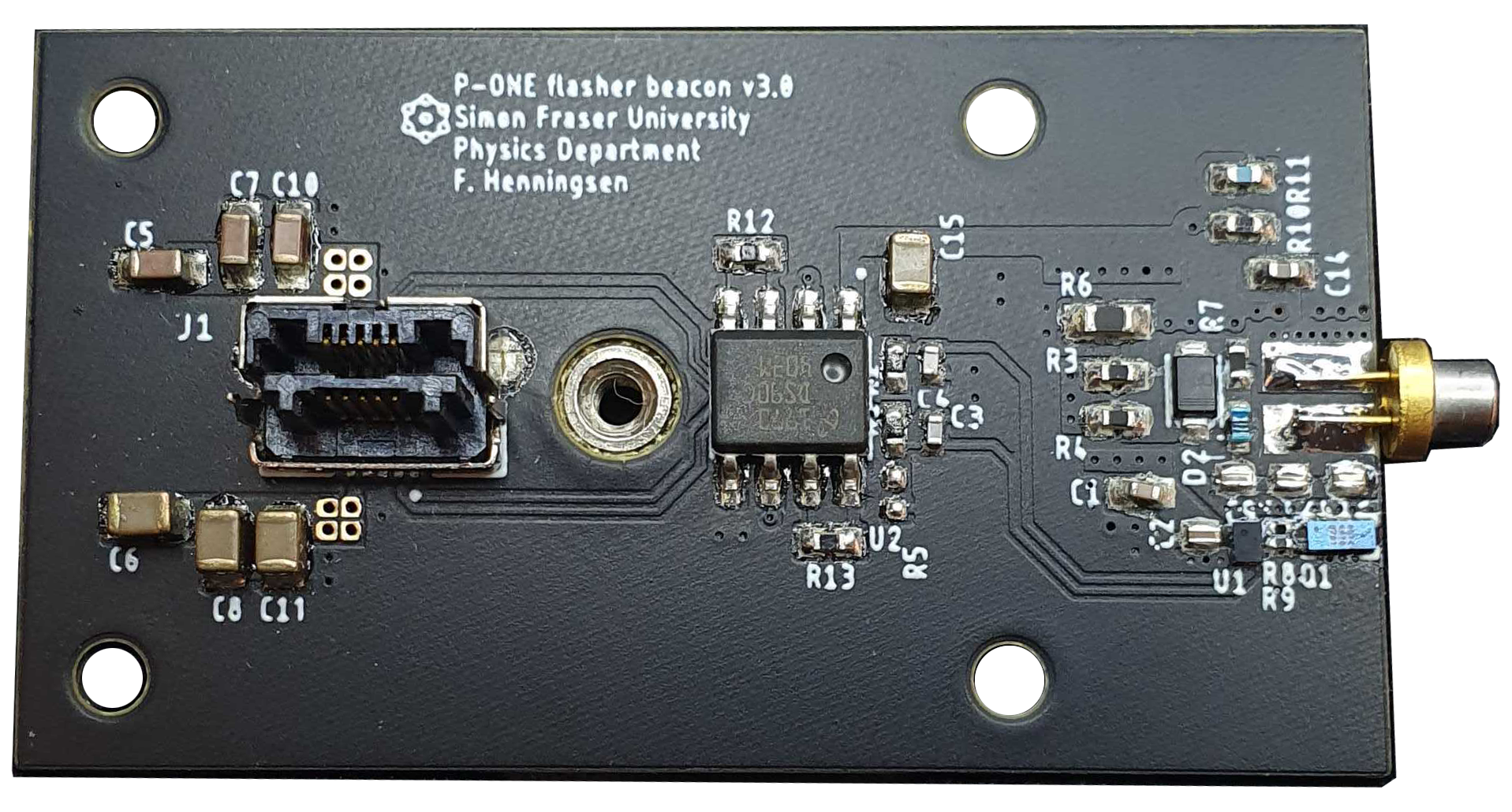}
      \caption{Assembled flasher PCB.}
      % \label{fig:setup-lab-b}
    \end{subfigure}
    \caption{\textbf{(a)} Baseline schematic of the GaNFET pulser circuit design showing core switching components, bias voltage, and AC-coupled trigger monitor for time-stamping (reproduced with permission from~\cite{henningsen_picosecond_2023,henningsen_self-monitoring_2020} and modified) and \textbf{(b)} a photograph of an assembled flasher PCB.}
    \label{fig:schematic-flasher}
\end{figure}

Light pulses are generated by providing a positive bias voltage and a gated trigger signal. The latter can be provided with either two dedicated trigger inputs or a single input and its delayed copy using a passive resistor-capacitor (RC) delay. In the P-ONE configuration, we use the latter to remove timing jitter between independent gate pulses. The minimum, stable driver pulse width supported by the LMG1020 is \unit[1]{ns}~\cite{texas_instruments_lmg1020_nodate}, but generated light pulses can reach sub-nanosecond timing with careful design of the circuit board layout and selection of components~\cite{henningsen_picosecond_2023}. Monitoring the cathode with an AC-coupled probing line also provides a direct response signal, allowing the trigger pulse arriving at the emitter diode to be time-stamped internally. A schematic of the driver circuit is shown in \cref{fig:schematic-flasher} together with an image of the assembled printed circuit board (PCB). 
\par%\noindent
Their main difference between LEDs and LDs is the emission mode and its characteristics; see~\cite{henningsen_picosecond_2023} and references therein for a detailed discussion. In short, LEDs emit light purely through spontaneous emission and are efficient at low current densities; LDs are less efficient in spontaneous emission mode but transition sharply to highly efficient stimulated emission above a threshold current. When lasing, LD light emission is typically fast, polarized, and shows a narrow spectral width. While fast pulses with a narrow spectrum are advantageous for timing-sensitive measurements like dispersion, the sharp intensity transition limits the achievable dynamic range. The emitter's physical size and placement on the PCB further determine the size of the discharge current loop and the inductance introduced into the driver system, both of which slow the pulse. In summary, careful selection of the emitter diode is required to meet the application's requirements.

As described above, the expected attenuation of the P-ONE water column requires fast, intense light pulses to meet calibration targets. For the directional flashers, we tested both LED and LD emitters, matching the central wavelengths critical for P-ONE. A total of $62$ LED and LD models were tested, and the final selection is detailed in \cref{tab:diodes-flashers}, including a summary of high-level performance metrics from testing 330 produced flasher units for integration into the first line.
\begin{table}[h!]
\centering
\begin{tabular*}{\textwidth}{l@{\extracolsep{\fill}} c c c c c c}
\toprule
\textbf{Diode} & $\mathbf{\lambda}$ & \textbf{Opening} $\theta_{1/2}$ & \textbf{Polarization} & \textbf{Photons} & \textbf{FWHM} & \textbf{Rise Time} \\
\cmidrule(r){2-2} \cmidrule(lr){3-3} \cmidrule(lr){4-4} \cmidrule(lr){5-5} \cmidrule(lr){6-7}
            & nm & deg & TE : TM   & per pulse          & \multicolumn{2}{c}{ns} \\[2pt]
\toprule
LED365-06Z  & $375$ & $4$            & -        & $10^{7} - 10^{9}$   & $3.4-4.0$ & $3.2-3.4$ \\
SLD3239VFR  & $405$ & $9\,/\,22$     & $50:1$   & $10^{8} - 10^{10}$  & $2.6-2.8$ & $3.0-3.0$ \\
PLPT5 447KA  & $447$ & $10\,/\,50$    & $100:1$  & $10^{8} - 10^{10}$  & $2.6-5.4$ & $2.8-3.6$ \\
GH0521DE2G  & $515$ & $8\,/\,22$     & $20:1$   & $10^{6} - 10^{10}$  & $5.8-1.4$ & $0.4-2.8$ \\
\bottomrule
\end{tabular*}
\caption{Emitter diode properties for directional flashers in P-ONE. Device characteristics are taken from manufacturer specifications. Opening angles are symmetric in both axes if one value is given and asymmetric otherwise. Note that measured performance metrics are the observed median values for 330 tested flasher units, given for bias voltages of \unit[$5$]{V} and \unit[$30$]{V}, respectively.}
\label{tab:diodes-flashers}
\end{table}\par\noindent
%

%%%%%%%%%%%%%%%%%%%%%%%%%%%%%%%%%%%%
% CHARACTERIZATION
%%%%%%%%%%%%%%%%%%%%%%%%%%%%%%%%%%%%
\subsection{Characterization}
Flasher units were tested with the measurement setup described in~\cite[][sec.~3]{henningsen_picosecond_2023}. This optical characterization leveraged multiple sensors to measure the light yield, time profile, and spectrum in parallel. Total systematic uncertainties of these measurements are estimated as \unit[15]{\%} on light yield, \unit[69]{ps} on timing, and \unit[6]{nm} on spectral measurements~\cite{henningsen_picosecond_2023}. This process also served as the Quality Control (QC) step, allowing flashers that did not meet specifications to be removed from the set. 

\begin{figure}[h!]
    \centering
    \vspace{-9pt}
    \begin{subfigure}{.95\textwidth}
      \centering
      \hspace*{0.2cm}
      \includegraphics[width=.99\textwidth]{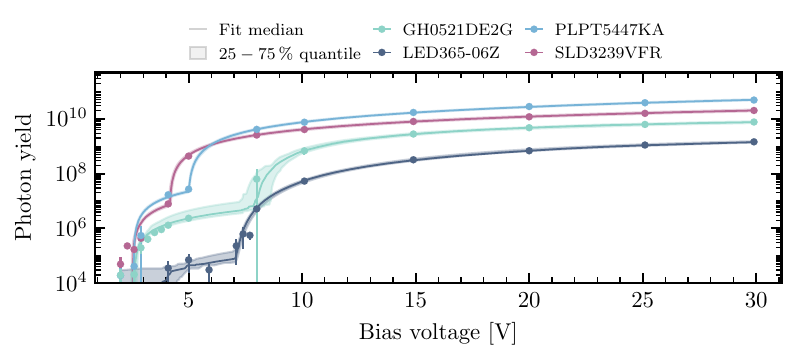}
    \end{subfigure}%
    % \vspace{2mm}
    \newline
    \begin{subfigure}{.99\textwidth}
      \centering
      \vspace{-3pt}
      \includegraphics[trim=0 0 0 1.1cm, clip, width=.96\textwidth]{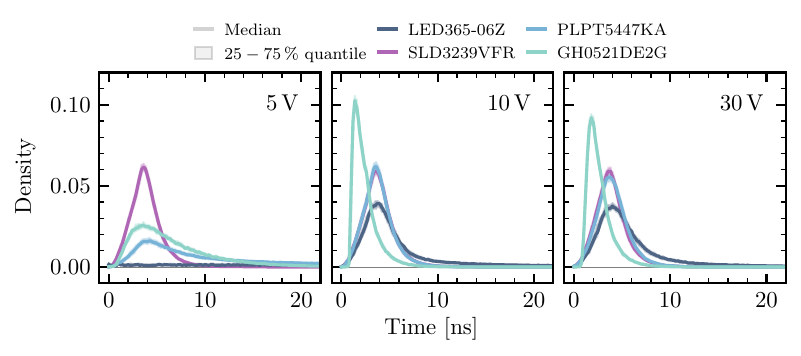}
    \end{subfigure}
    \newline
    \begin{subfigure}{.99\textwidth}
      \centering
      \vspace{-3pt}
      \includegraphics[trim=0 0 0 1.1cm, clip, width=.99\textwidth]{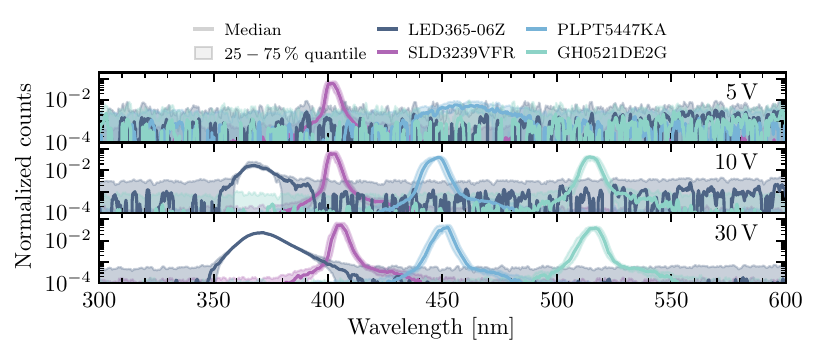}
    \end{subfigure}
    \caption{Photon yield (top), time profile (center) and spectrum (bottom) results for all directional flasher units that passed QC. Lines and shaded bands show the median and \unit[$25-75$]{\%} quantiles, respectively. For the photon yield, the fit combines a linear component with a reverse-sigmoid activation function, intended to model the lasing onset. For the time profile and spectrum, the continuous lines are obtained by linear interpolation of raw timing and spectral data, binned in \unit[20]{ps} and \unit[2]{nm} intervals, respectively. Top- and center-row figures are partly reproduced from~\cite{ghuman_situ_2025}.}
    \label{fig:flasher-summary}
\end{figure}

A total of 330 produced flashers were characterized in this setup for the first P-ONE line. Measurements on each flasher unit are carried out over a bias voltage range of \unit[$0-30$]{V} and frequencies between \unit[100]{Hz} and \unit[100]{kHz}. Time profiles and spectra are measured at select voltages to optimize the total characterization time per unit. Twelve of the tested units failed QC. Measurement results for the full population of production flashers are presented in \cref{fig:flasher-summary}. In general, the flashers show remarkable reproducibility in all measured metrics. Once deployed, the directional flashers are expected to operate at a bias voltage above \unit[$10$]{V}, at which all channels meet the requirements.
\par%\noindent
A subset of flashers underwent longevity testing to evaluate endurance during extended operation. This assessment is essential for estimating the functional lifetime of individual units, particularly in the context of mooring lines designed for operating lifetimes of up to \unit[20]{years}. 

\begin{figure}[h!]
    \centering
    % \vspace{-12pt}
    \includegraphics[width=.9\textwidth]{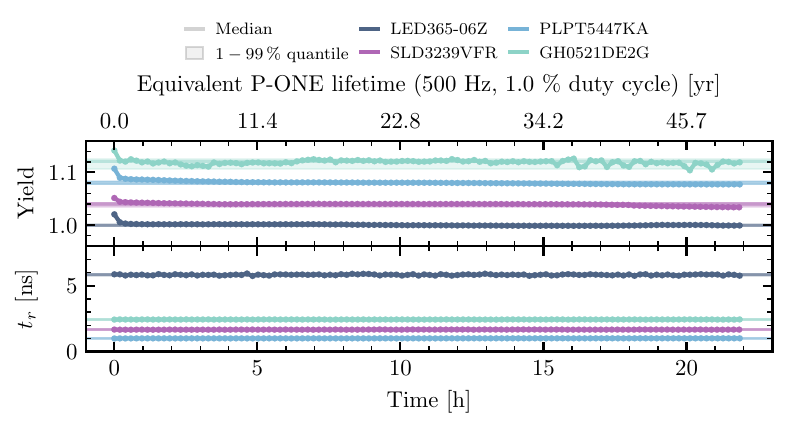}
    \caption{Aging test results for one of each flasher board type. Shown are the relative light yield and rise time of the observing SiPM for flasher pulses as a function of total operating time. Yield medians across different channels have been artificially offset to improve visual separation. Flashers were operated at a bias voltage of \unit[24]{V} and a frequency of \unit[100]{kHz}. The corresponding P-ONE equivalent lifetime is shown as a second axis.}
    \label{fig:flasher-aging}
\end{figure}

To simulate accelerated aging in a reasonable time frame, a flasher unit is installed into the measurement setup and operated continuously at a frequency of \unit[100]{kHz} and a bias voltage of \unit[24]{V}. Given the total number of emitted pulses during this time, a 12-hour measurement period corresponds to an equivalent P-ONE lifetime of more than \unit[27]{years} (assuming a P-ONE pulsing frequency of \unit[500]{Hz} with a duty cycle of \unit[1]{\%}). The results of these measurements are shown in \cref{fig:flasher-aging}. All units remained functional and stable during the longevity test and exhibited controlled burn-in, with at most \unit[$1-2$]{\%} degradation over the measurement time.
\par%\noindent
%

%%%%%%%%%%%%%%%%%%%%%%%%%%%%%%%%%%%%
% INTEGRATION
%%%%%%%%%%%%%%%%%%%%%%%%%%%%%%%%%%%%
\subsection{Integration}
The P-ONE module assembly is based on a 3D-printed mounting frame that houses all components and is mounted to the Titanium flange of the pressure housing. In every hemisphere, this defines the location of eight PMTs and five cut-out locations (see \cref{fig:sketch-flashers}). The cut-outs are meant to allow a generic mechanical interface for auxiliary instrumentation. For the first line, they are intended for acoustic receivers for positioning~\cite{agostini_prototype_2025} and the directional flashers described here.

\begin{figure}[h!]
    \centering
    % \vspace{-12pt}
    \includegraphics[width=.9\textwidth]{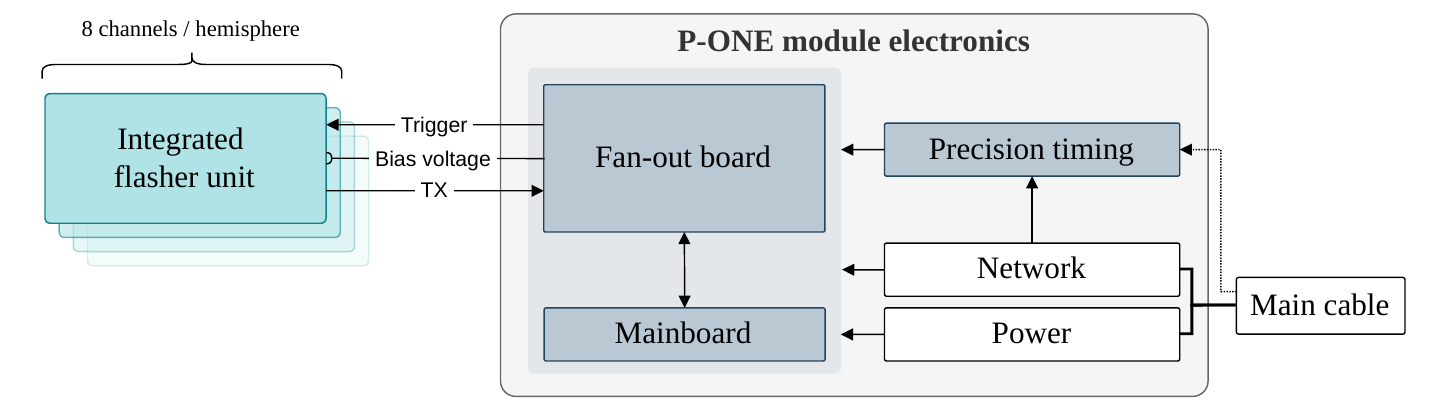}
    \caption{Schematic flowchart of the flasher unit integration into P-ONE module electronics.}
    \label{fig:flow-flasher}
    \vspace{-12pt}
\end{figure}

The P-ONE module electronics (see \cref{fig:flow-flasher}) include a mainboard connected to the main cable and a fan-out board in each hemisphere. The main cable connection provides fiber-optic network connectivity and precision timing; the fan-out boards connect to the mainboard and link all devices integrated into the respective hemispheres. For the flashers, the fan-out board provides the bias voltages and forwards the mainboard trigger signals. The flasher feedback is fed into a discriminator on the fan-out board and time-stamped with a time-to-digital converter (TDC) on the mainboard.
\par
\begin{figure}[h!]
    \centering
    \begin{subfigure}{.32\textwidth}
      \centering
      \includegraphics[width=.99\textwidth]{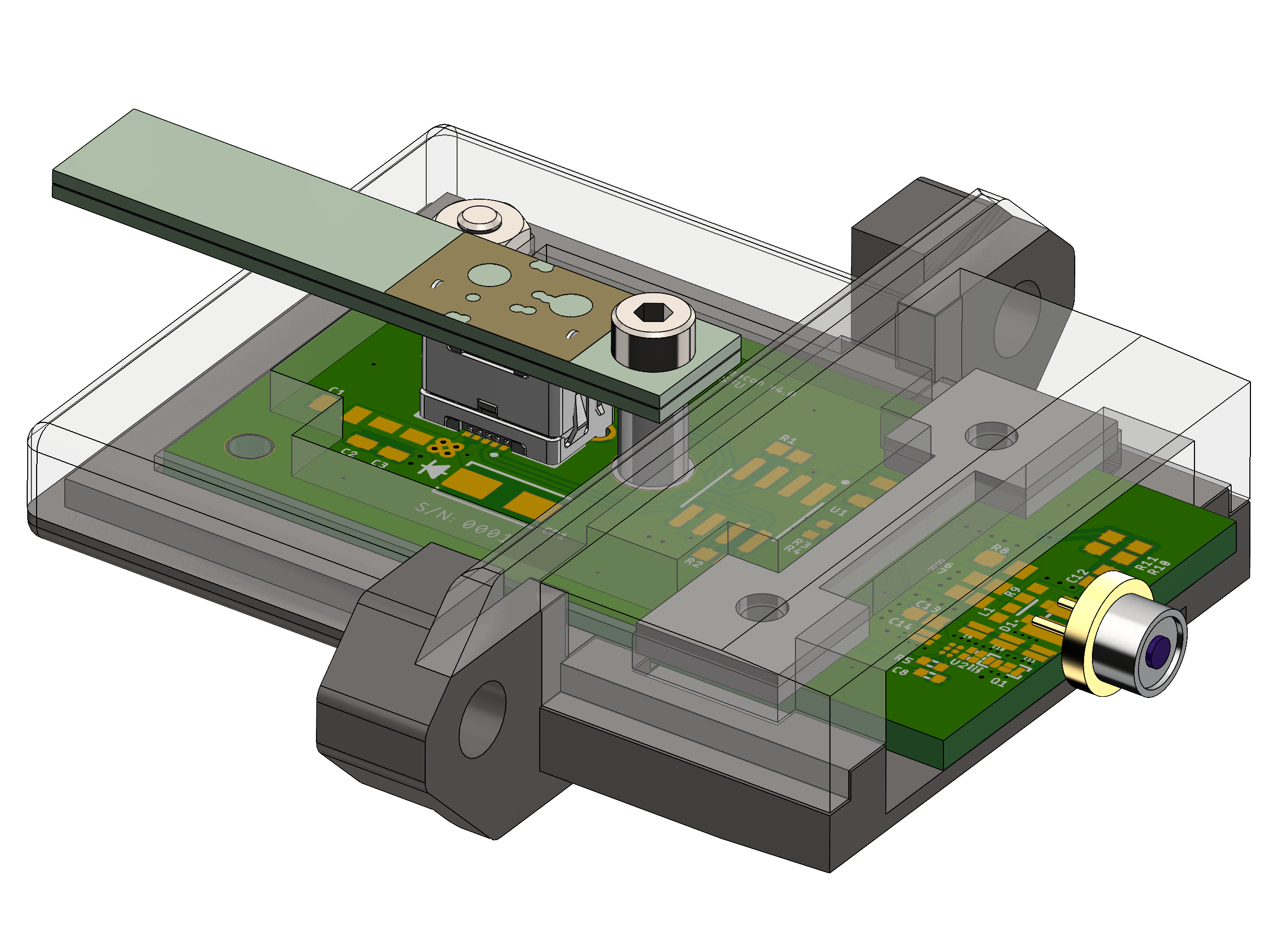}
      \caption{Flasher bracket with gasket.}
      \label{fig:mechanical-flasher-a}
    \end{subfigure}%
    \hspace{1.5mm}
    \begin{subfigure}{.32\textwidth}
      \centering
      \includegraphics[width=.99\textwidth]{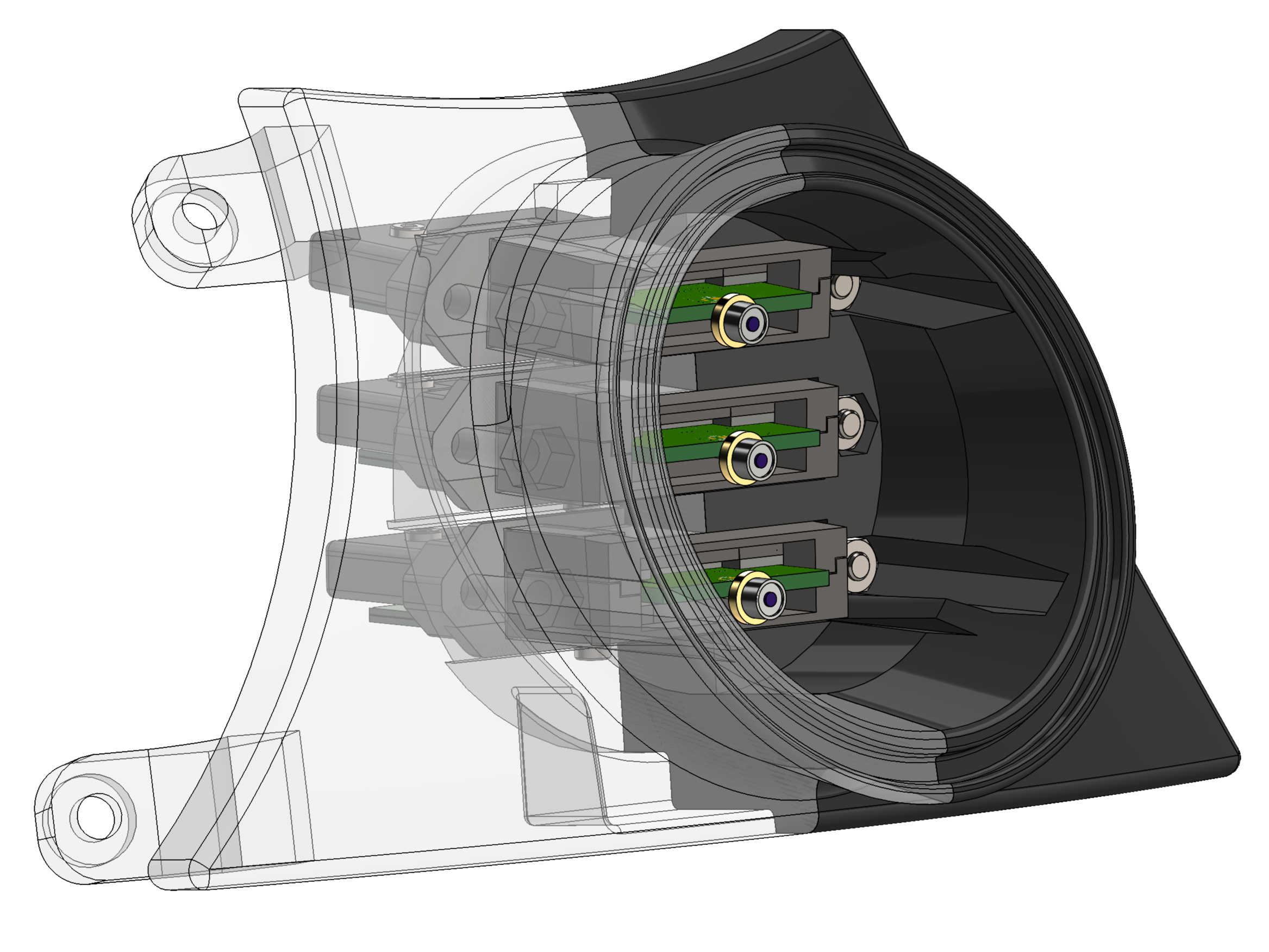}
      \caption{Up/down unit with brackets.}
      \label{fig:mechanical-flasher-b}
    \end{subfigure}
    \hspace{1.5mm}
    \begin{subfigure}{.32\textwidth}
      \centering
      \includegraphics[width=.99\textwidth]{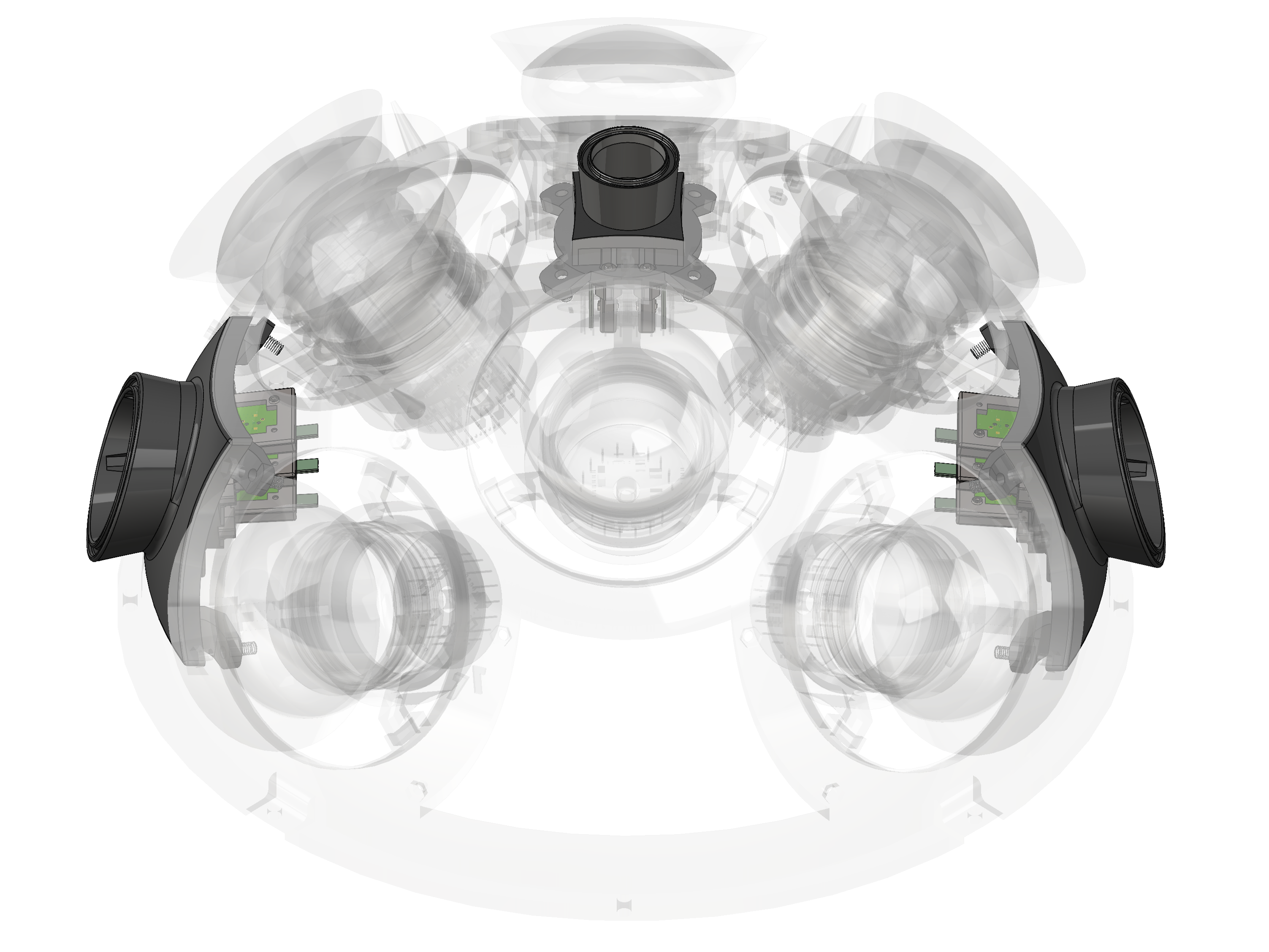}
      \caption{Integrated units in hemisphere.}
      \label{fig:mechanical-flasher-c}
    \end{subfigure}
    \caption{Mechanical integration of directional flashers from \textbf{(a)} bracket (taken from~\cite{ghuman_situ_2025}) to \textbf{(b)} assembled units and finally \textbf{(c)} integration into the mounting frame within a hemisphere.}
    \label{fig:mechanical-flasher}
    % \vspace{-\baselineskip}
\end{figure}

%\noindent
As shown in \cref{fig:mechanical-flasher}, assembled flasher PCBs with mounted emitters are mechanically integrated into P-ONE modules in three stages: First, a mounting bracket is assembled to encase the PCB and provide an optical seal to the emitter section on the board. Then, these brackets are mounted to 3D-printed parts, which define flasher orientation and integrate into the cut-out positions in the mounting frame. Finally, these units are equipped with rubber o-rings and mounted to the mounting frames with spring-loaded screws. Rubber gaskets and O-rings are used to minimize light leakage.

The light from upward- and downward-facing flashers exhibits angular dispersion and is not emitted perpendicular to the curvature of the glass. The varying refractive indices between the air-glass-water interfaces produces a deviation from the vertical. To correct for this, the mechanical orientation of each flasher was optimized using simulations in GEANT4~\cite{agostinelli_geant4_2003}. This aligns the central emission axis, per flasher position and wavelength, with the detection line. Incorporating emission profiles, wavelengths, positions, and the expected seawater refractive index~\cite{schuster_measurement_2002}, we find that the required correction angle increases with distance from the glass equator (referring to \cref{fig:sketch-flashers}) with tilts of about \unit[$4-6$]{degrees}. A photo of a directional flasher assembly is shown in \cref{fig:photos-a}.
\par%\noindent
Three special positions along the line are equipped with directional axicon flashers that generate Cherenkov ring–like light patterns for calibration studies of timing and optical scattering. Each unit integrates an LD-450-3000G laser diode by Roithner Lasertechnik driven by the standard flasher electronics. One axicon points upward, one downward (both with a \unit[20]{degree} aperture), and one perpendicular to the line (\unit[$\approx 60$]{degree} aperture). The upward- and downward-facing units enable measurements of time synchronization, photomultiplier efficiency, and sedimentation effects within the illuminated cone. The primary purpose of all axicons is the determination of long-distance water optical properties through measurements of light scattering inside and outside the cone. The sideways-oriented axicon additionally allows for future inter-string and geometry measurements.

%%%% P-CAL %%%%%%%%%%%
\section{Optical calibration module}
\label{sec:pcal}
\begin{figure}[h!]
    \centering
    \begin{subfigure}{.49\textwidth}
      \centering
      \includegraphics[width=.85\textwidth]{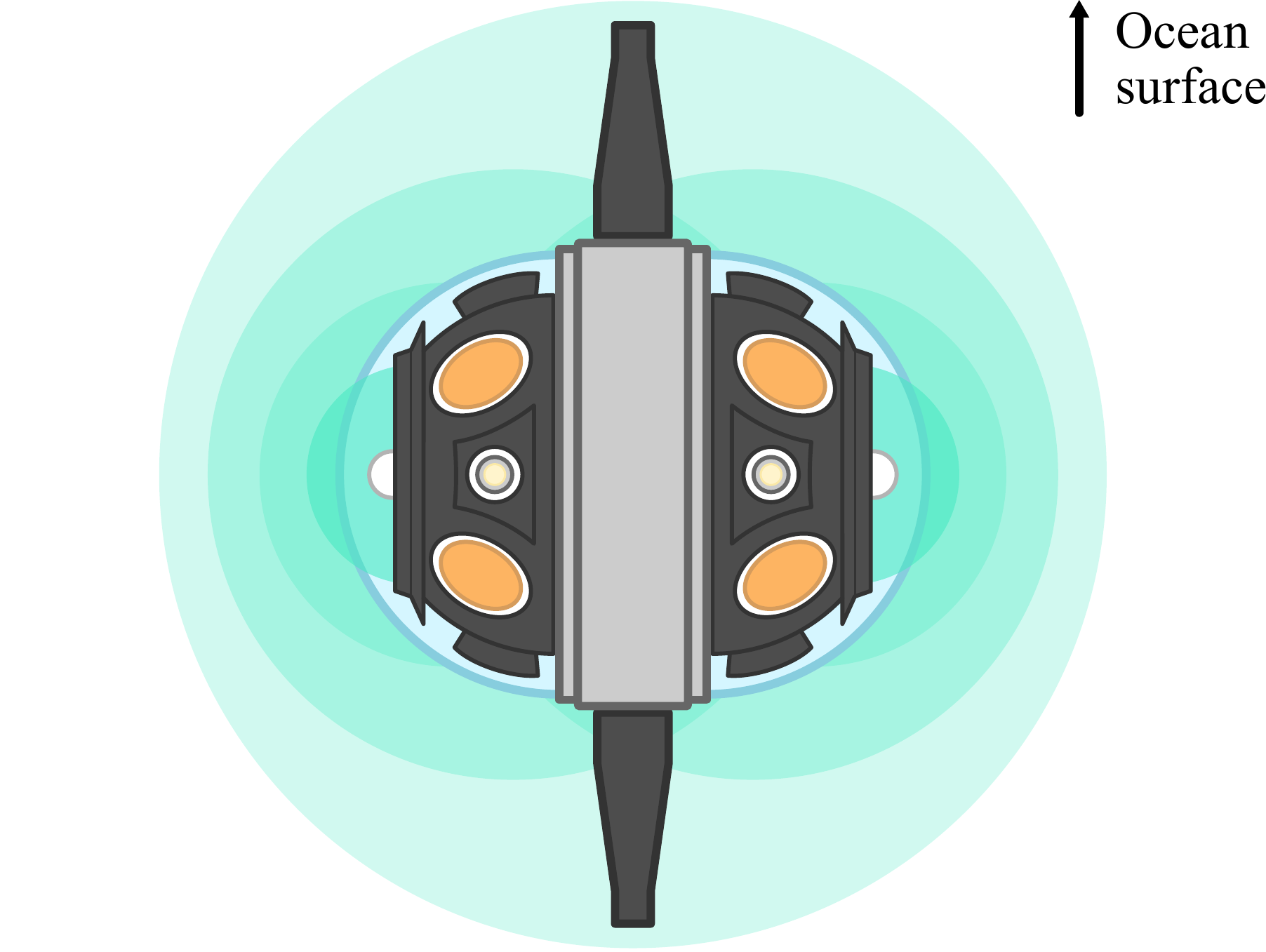}
      \caption{Isotropic emission from both hemispheres.}
      % \label{fig:setup-lab-a}
    \end{subfigure}%
    \hspace{1mm}
    \begin{subfigure}{.49\textwidth}
      \centering
      \includegraphics[width=.85\textwidth]{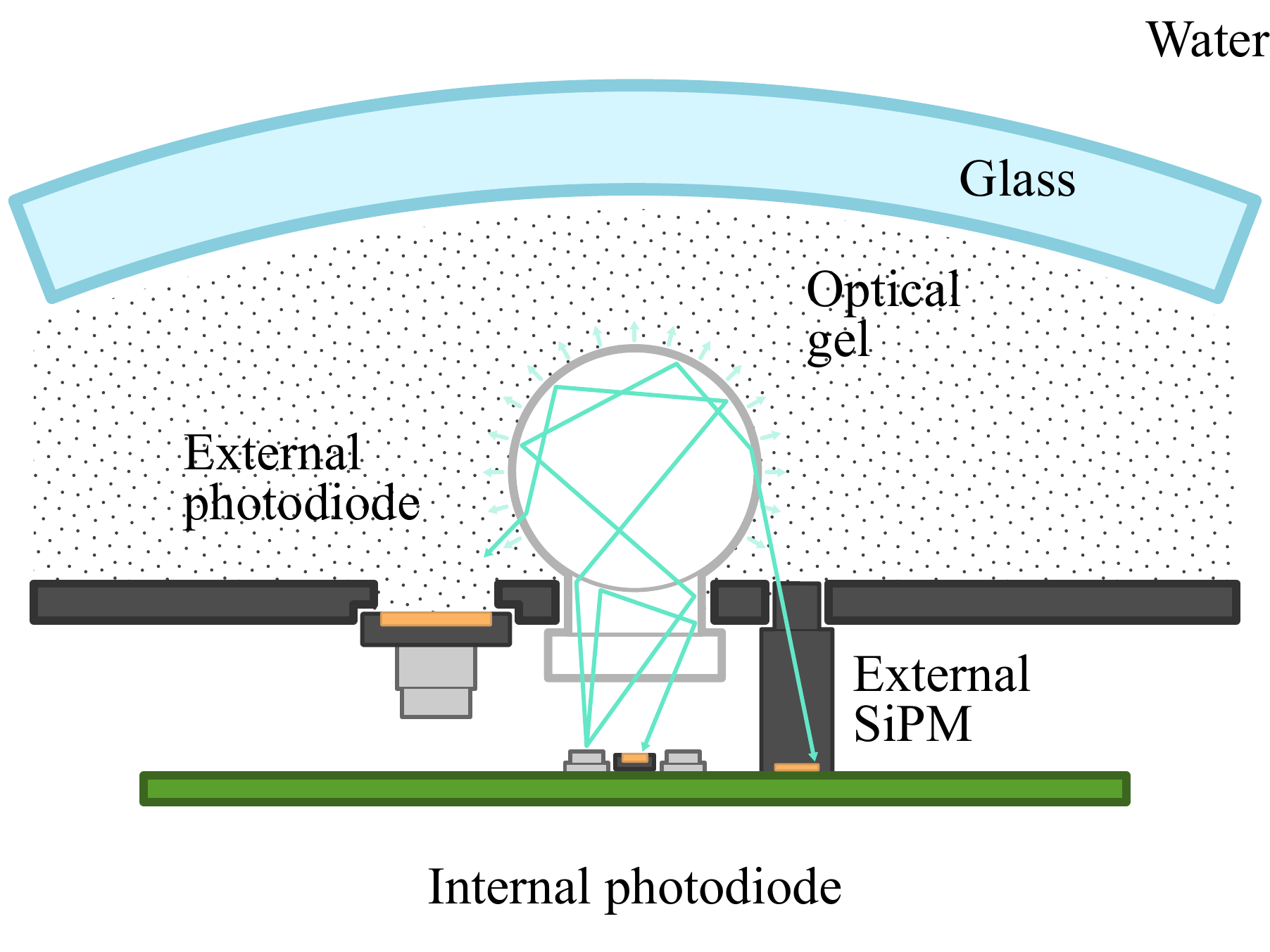}
      \caption{P-CAL diffuser and self-monitoring systems.}
      % \label{fig:setup-lab-b}
    \end{subfigure}
    \caption{Visualizations of the P-CAL in P-ONE. Shown are \textbf{(a)} the full, isotropic emission and \textbf{(b)} a detailed view of the isotropic flasher and self-monitoring concept. Both adapted from~\cite{ghuman_situ_2025}.}
    \label{fig:sketch-pcal}
\end{figure}

The large, instrumented water body of P-ONE requires comprehensive calibration throughout the detection volume. To meet this need, we developed the hybrid P-CAL instrument, which combines a multi-wavelength, isotropic light source with self-monitoring sensors and photomultiplier tubes (PMTs). A single isotropic light pulse can illuminate many sensors in the three-dimensional array, minimizing directional uncertainties in instrument alignment and reducing the impact of the light source's intrinsic uncertainties. This enhances the overall reliability and precision of the calibration. The design of the isotropic light emission and self-monitoring strategy builds on a concept originally developed for an IceCube Upgrade instrument~\cite{henningsen_self-monitoring_2020,henningsen_self-monitoring_2021} and adapted for deployment in P-ONE. In addition to the energy scale (i.e., water properties and module efficiencies), the P-CAL will facilitate optical calibration of detector geometry by measuring inter-module distances. Two P-CAL modules will be installed on the first line, and the core concept is shown in \cref{fig:sketch-pcal}.
\par%\noindent

\subsection{Requirements}
The main objective for the P-CAL is the optical calibration of the water column and therefore the detector's energy scale. By emitting isotropic light pulses into the multi-line array, volumetric measurements enable detailed calibration campaigns of optical water properties and module efficiencies. For this, the P-CAL comprises three main components: the Calibration Board (CB), which combines light pulsing and photosensor circuitry for in situ self-monitoring of emitted pulses; an optical diffuser design that creates isotropic light pulses when combining two hemispheres; and a camera system that monitors the glass pressure housing for sedimentation and biofouling.

A first requirement for the instrument is to use the same light-pulsing approach as the directional flashers to unify the internal electronics design. That means individual light-pulsing circuits also use the GaNFET concept described in \cref{sec:directional}, and diodes are restricted to the transparency window in the water column, \unit[$350-550$]{nm}, to achieve sufficient optical reach. The volumetric spread of the light pulses and their $r^{-2}$ dependence requires a higher light yield of at least \unit[$10^{9}-10^{10}$]{photons per pulse} to reach a minimum of two modules (or \unit[100]{m}) with sufficient photon statistics. We thus allow slightly longer pulses with a FWHM of less than \unit[5]{ns}. Secondly, the light-pulsing circuits are complemented by photosensor monitoring channels for in situ self-monitoring of all emitted pulses. Together with laboratory calibration prior to deployment, this enables monitoring of the per-pulse characteristics of the deployed modules.

\subsection{Isotropic emitter design}
\begin{figure}[h!]
    \centering
    \includegraphics[width=.99\textwidth]{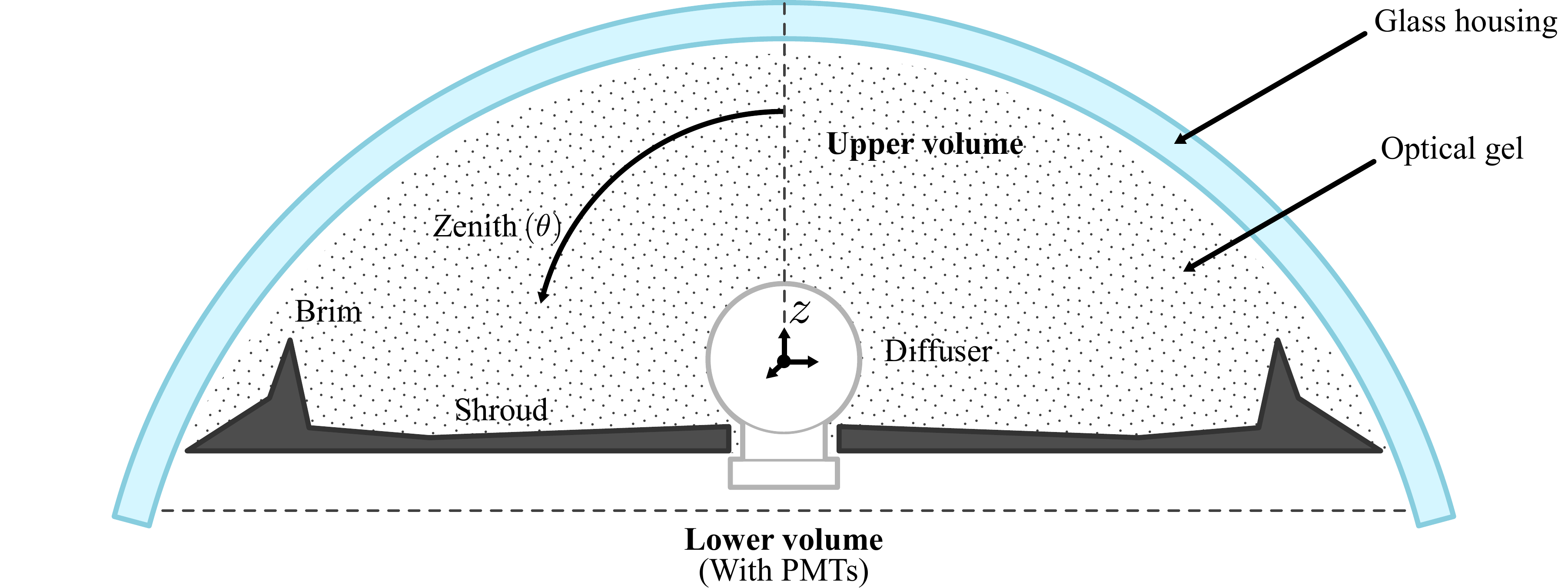}
    \caption{Simplified geometry of the emission section in the P-CAL. Shown are the glass pressure housing, the diffuser, and the shroud with its brim, which divides the module into upper and lower volumes. The upper volume is filled with optical gel to compensate refraction effects in water.}
    \label{fig:pcal-geometry}
\end{figure}

The design of the P-CAL emission system is based on a concept developed for the IceCube Upgrade~\cite{henningsen_self-monitoring_2020,henningsen_self-monitoring_2021}, which uses optical polytetrafluoroethylene (PTFE) shaped into a hollow, semi-transparent integrating sphere. A PTFE plug complementing the sphere provides a first stage of diffusion for the anisotropic light pulse emitted from the diodes. As pictured in \cref{fig:sketch-pcal}, the hybrid nature of the instrument means that four PMTs in each P-OM hemisphere are replaced with the optical diffuser and related components. The diffuser is positioned off-center relative to the glass hemisphere's z-axis, and careful optical design is required to achieve isotropic emission. A planar shroud with a circular brim divides the inner volume into a lower volume containing PMTs and an upper volume for light emission. The diffuser, its mechanical stack, and the electronics board hosting flasher circuitry are assembled together first and subsequently fixed to the shroud. Optical gel in the emission volume compensates for refraction and lensing effects in water. A simplified cross-section sketch of this geometry and its major components is shown in \cref{fig:pcal-geometry}.
\par
\begin{figure}[h!]
    \centering
    \includegraphics[width=.99\textwidth]{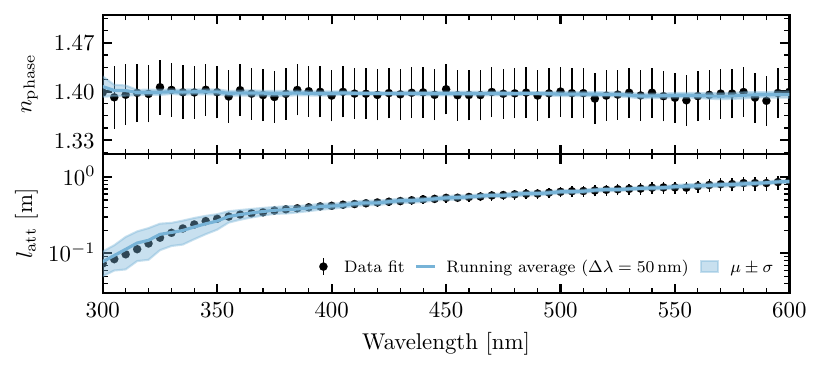}
    \caption{Phase refractive index (top) and attenuation length (bottom) of Wacker SilGel 612 (mixed with a ratio of 1.5:1), obtained from fits to several transmission measurements of gel slabs.}
    \label{fig:pcal-gel-refindex}
\end{figure}

%\noindent
The optical design is optimized for isotropic emission using a few key geometric parameters: the lateral position of the diffuser center, the position and dimension of the shroud brim near the glass, and the refractive index of the medium in the upper volume. Mirroring the procedure used for the directional flashers, we use a GEANT4 simulation to optimize the design, along with laboratory measurements. These simulations confirm that an optical gel with a refractive index similar to water can accommodate an off-axis diffuser by reducing asymmetric refraction and lensing effects. We therefore use two-component Wacker SilGel 612 with a mixing ratio of 1.5:1 and an approximate refractive index of $1.404$ at \unit[532]{nm}~\cite{noauthor_wacker_2014}. We further measured wavelength-dependent transmission of the gel using the Agilent 60 UV-Vis spectrophotometer. The refractive index and attenuation length were obtained through a combined fit to the measured transmission curves. Results are shown in \cref{fig:pcal-gel-refindex}, confirming an effectively constant refractive index between \unit[$300-600$]{nm}.
\par
\begin{figure}[h!]
    \centering
    \includegraphics[width=.99\textwidth]{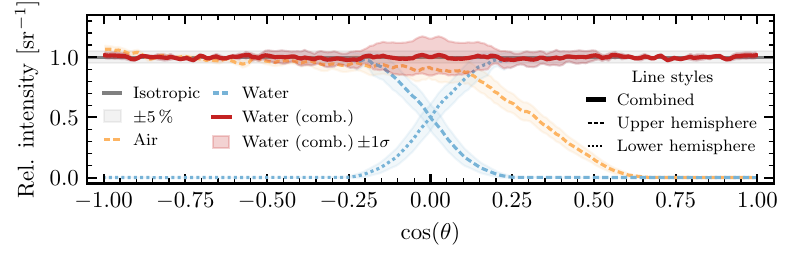}
    \caption{GEANT4 simulations of optimized P-CAL emission intensities versus zenith angle at \unit[2.5]{m} distance in air and water. Shown are the individual hemisphere contributions and the combined profile in water. The indicated error bands show the standard deviations from an ensemble of simulated parameter realizations within design uncertainties.}
    \label{fig:pcal-geant4}
\end{figure}

\noindent
The optical design is optimized using a P-CAL geometry implementation in GEANT4. In combination with in-air measurements (see \cref{sec:pcal:characterization}), simulated surface properties of the various components were determined and the optical properties of the gel were set to their measured values. To optimize the P-CAL emission profile for isotropy in P-ONE, seawater is optically characterized as given in~\cite{smith_optical_1981,schuster_measurement_2002}. The resulting simulated emission profiles are shown in \cref{fig:pcal-geant4} for both air and ocean water as the external medium, and normalized to the average of $\cos(\theta) \in [-1, -0.5]$.
\par\noindent

\subsection{Light pulser and self-monitoring design}
\label{sec:pcal:cb}
The Calibration Board (CB) hosts both light pulsers and self-monitoring sensor circuitry. The light pulsers use the circuit design of the directional flashers (see \cref{sec:directional}) for a total of five light pulser channels in each hemisphere. The three default channels use the high-power design described in~\cite{henningsen_picosecond_2023} and are optimized for intensity, with target pulse widths of about \unit[3]{ns} using the EPC2214 GaNFET; the two remaining channels use the faster directional-flash design and are optimized for timing. The pulser configurations and selected emitter diodes for the P-CAL, along with their performance characteristics, are given in \cref{tab:diodes-pcal}. It is important to note that the opening angles and polarizations are not specified here, as the diffuser effectively randomizes the output light.
\begin{table}[h!]
\centering
\begin{tabular*}{\textwidth}{l@{\extracolsep{\fill}} c c c c c c}
\toprule
\textbf{Diode}    & $\mathbf{\lambda}$ & \textbf{Mode} & $\mathbf{V_\mathrm{max}}$ & \textbf{Photons} & \textbf{FWHM} & \textbf{Rise Time} \\
\cmidrule(r){2-2} \cmidrule(lr){3-3} \cmidrule(lr){4-4} \cmidrule(lr){5-5} \cmidrule(lr){6-7}
                  & nm      &  -     & V & per pulse            & \multicolumn{2}{c}{ns} \\[2pt]
\toprule
SLD3239VFR        &  $405$  &  fast  &  $30$  &  $10^{8} - 10^{10}$   &  $1.5-1.5$  &  $0.5-0.9$ \\
RLT405-600MGE     &  $405$  &  -     &  $30$  &  $10^{9} - 10^{11}$  &  $2.9-3.4$  &  $3.1-3.1$ \\
LD-450-3000G      &  $450$  &  fast  &  $30$  &  $10^{7} - 10^{10}$  &  $11.2-2.6$  &  $3.4-1.7$ \\
RLT455-5W-GOP-FAC &  $455$  &  -     &  $60$  &  $10^{8} - 10^{11}$  &  $3.1-2.8$  &  $2.0-1.7$ \\
NLD521000G        &  $510$  &  -     &  $60$  &  $10^{8} - 10^{11}$  &  $6.4-2.3$  &  $4.1-1.6$ \\
\bottomrule
\end{tabular*}
\caption{Emitter diode properties for the P-CAL flashers in P-ONE. Device characteristics are taken from manufacturer specifications. Note that measured performance metrics are the observed median values for 40 tested channels, given for bias voltages of \unit[$5$]{V} and \unit[$30$]{V}, respectively.}
\label{tab:diodes-pcal}
\end{table}\par
\begin{figure}[h!]
    \vspace{-6pt}
    \centering
    \begin{subfigure}{.49\textwidth}
      \centering
      \includegraphics[width=.85\textwidth]{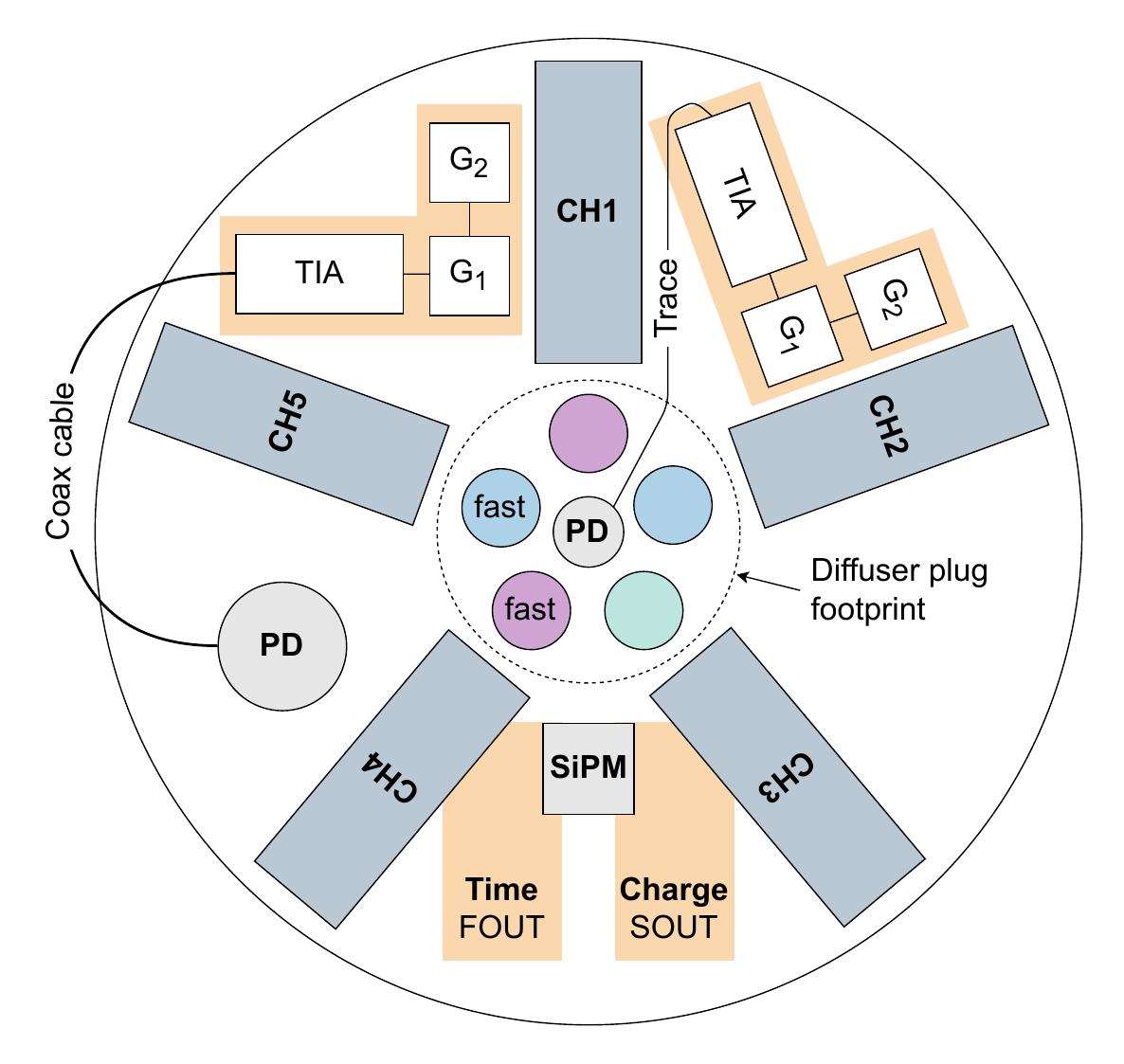}
      \caption{Conceptual Calibration Board layout.}
      % \label{fig:setup-lab-a}
    \end{subfigure}
    \begin{subfigure}{.49\textwidth}
      \centering
      \includegraphics[width=.80\textwidth]{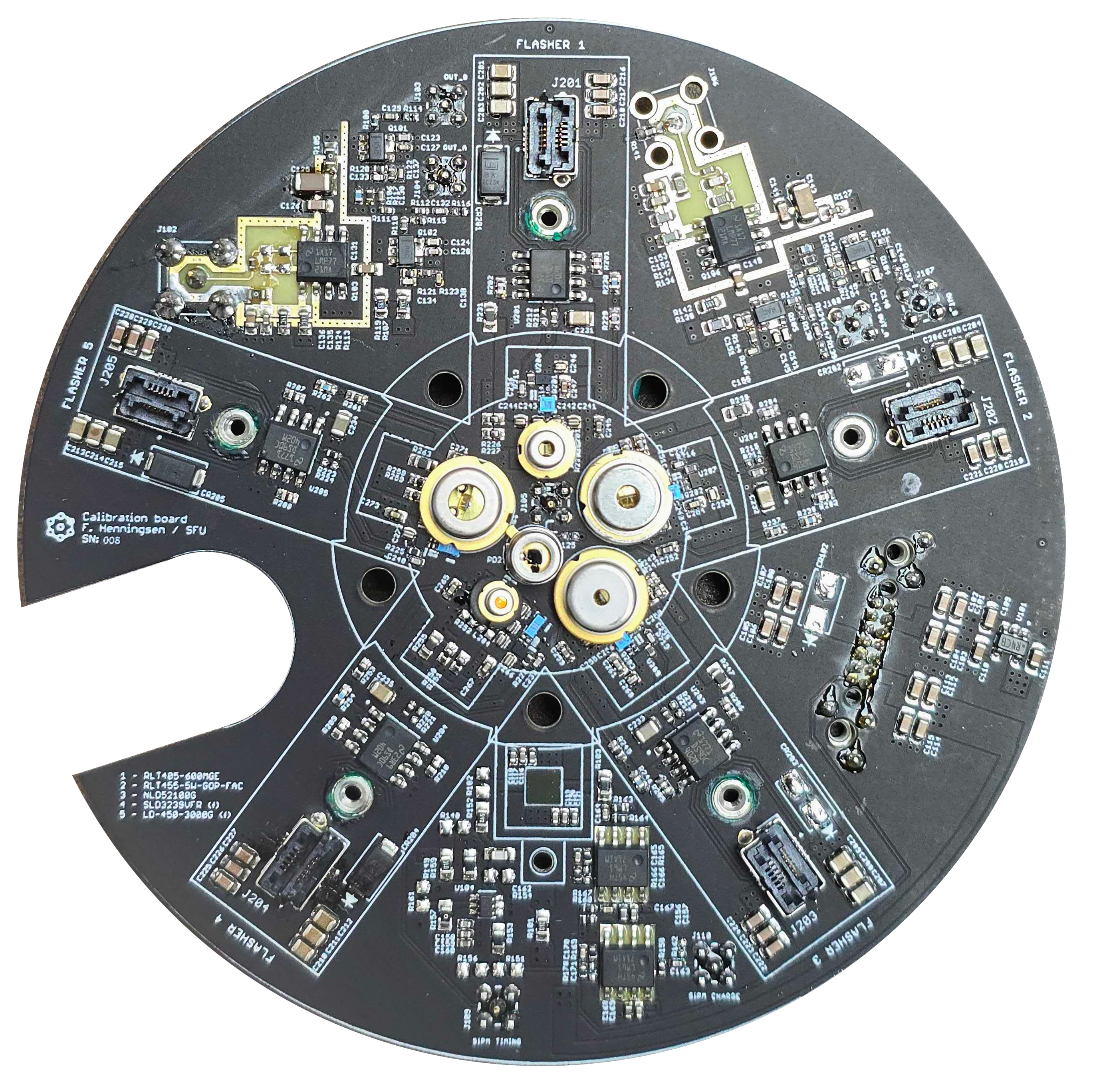}
      \caption{Assembled Calibration Board.}
      % \label{fig:setup-lab-b}
    \end{subfigure}
    \caption{Concept and layout of the P-CAL Calibration Board. The images show \textbf{(a)} the conceptual layout with flashers and self-monitoring senors and \textbf{(b)} a photo of a fully-assembled PCB.}
    \label{fig:sketch-cb}
    \vspace{-\baselineskip}
\end{figure}

\noindent
The self-monitoring sensors are the Hamamatsu S2281-01~\cite{hamamatsu_si_2015} and S1226-18BK~\cite{hamamatsu_si_2023} photodiodes and the onsemi MICROFJ-30035 Silicon Photomultiplier (SiPM)~\cite{onsemi_jseries_2017}. The small S1226 photodiode sits just under the diffuser plug, acting as an internal sensor; the others measure the light output through openings in the shroud. This concept and an assembled board are shown in \cref{fig:sketch-cb}.
\par
\begin{figure}[h!]
    \centering
    \vspace{-6pt}
    \begin{subfigure}{.98\textwidth}
      \centering
      \includegraphics[width=.75\textwidth]{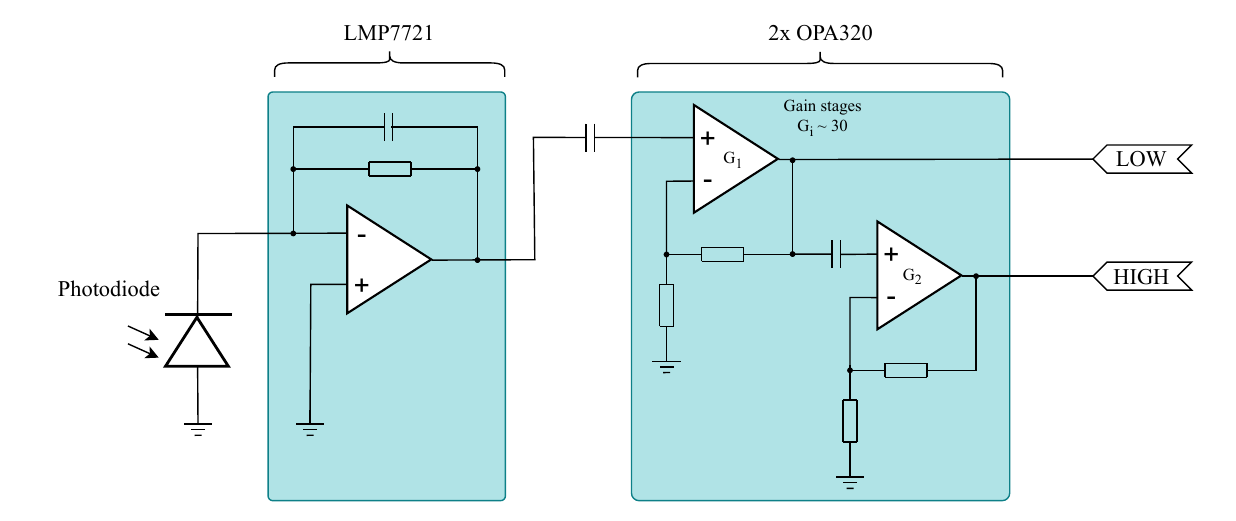}
      \caption{Photodiode trans-impedance amplifier with subsequent voltage amplifiers.}
    \end{subfigure}%
    % \vspace{2mm}
    \newline
    \begin{subfigure}{.49\textwidth}
      \centering
      \includegraphics[width=.8\textwidth]{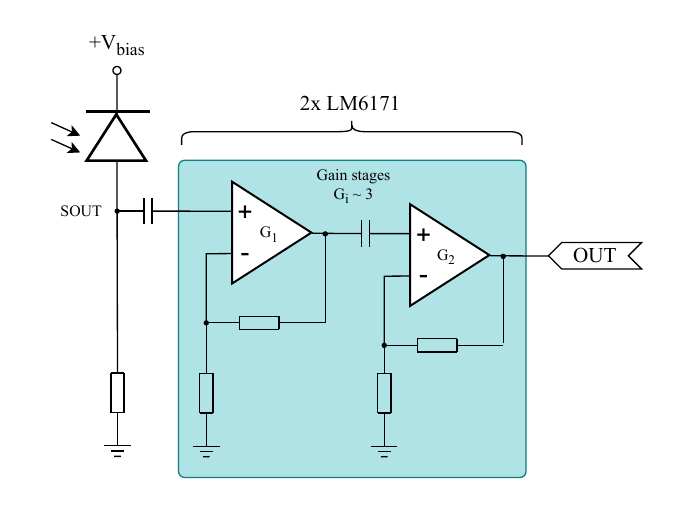}
      \caption{SiPM voltage amplifier.}
    \end{subfigure}
    % \hspace{1mm}
    \begin{subfigure}{.49\textwidth}
      \centering
      \includegraphics[width=.8\textwidth]{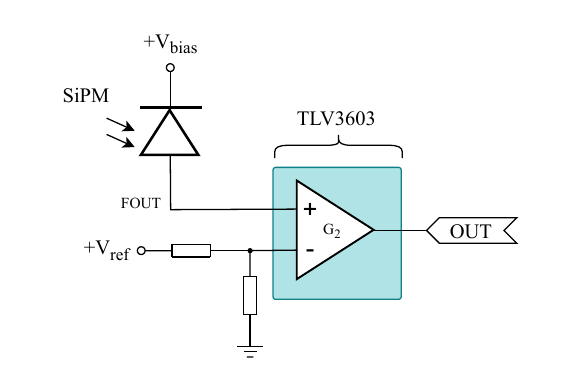}
      \caption{SiPM timing discriminator.}
    \end{subfigure}
    \caption{Conceptual schematics of the \textbf{(a)} trans-impedance and voltage amplifiers used for the charge measurement of both photodiodes, as well as the SiPM \textbf{(b)} voltage amplifier and  \textbf{(c)} timing discriminator for the charge and timing readout through its SOUT and FOUT pins, respectively.}
    \label{fig:pcal-cb-schematics}
\end{figure}

%\noindent
All sensor signals are shaped by preamplifier circuitry on the CB and digitized on the mainboard. For each photodiode, we use the LMP7721 in a transimpedance amplifier (TIA) configuration, followed by two stages of OPA320 voltage amplifiers, both from Texas Instruments. The TIA in the photodiode charge readout channel converts the photocurrent from each light pulse into a positive voltage pulse. The subsequent voltage amplifier adds a gain of about $30$ to the low-gain output stage and an additional $30$ to the high-gain output stage. This circuitry is applied to each photodiode, providing a total of four output channels for photodiode charge measurements per hemisphere. The SiPM has two output pins: a standard (SOUT) and a fast (FOUT) output. The latter is an AC-coupled channel that is used for timing and is connected to the Texas Instruments TLV3603 timing discriminator. This uses a small tripping voltage to provide rectangular output signals for the pulse start time. The SOUT pin is connected to the Texas Instruments LM6171 voltage amplifier, and the signal gain can be tuned by adjusting the intrinsic SiPM gain via its applied reverse bias voltage. All circuit schematics are shown conceptually in \cref{fig:pcal-cb-schematics}.
\par\noindent

\subsection{Characterization}
\label{sec:pcal:characterization}
A total of eight CBs were produced and assembled, totaling 40 flasher and 32 self-monitoring sensor channels, respectively. QC measurements and characterizations were performed for all CBs. 

\begin{figure}[h!]
    \centering
    \vspace{-12pt}
    \begin{subfigure}{.94\textwidth}
      \centering
      \hspace*{0.2cm}
      \includegraphics[width=0.99\textwidth]{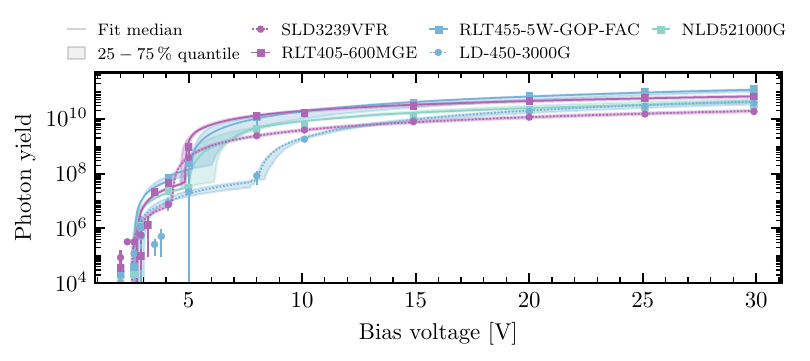}
    \end{subfigure}%
    % \vspace{2mm}
    \newline
    \begin{subfigure}{.99\textwidth}
      \centering
      \vspace{-3pt}
      \includegraphics[trim=0 0 0 1.1cm, clip, width=.95\textwidth]{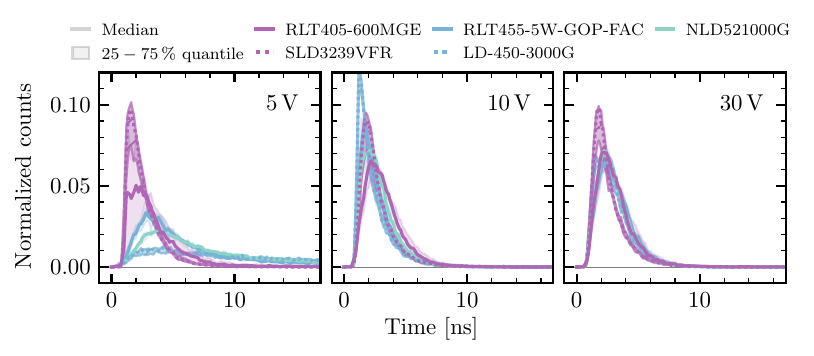}
    \end{subfigure}
    \newline
    \begin{subfigure}{.95\textwidth}
      \centering
      \vspace{-3pt}
      \includegraphics[trim=0 0 0 1.1cm, clip, width=.99\textwidth]{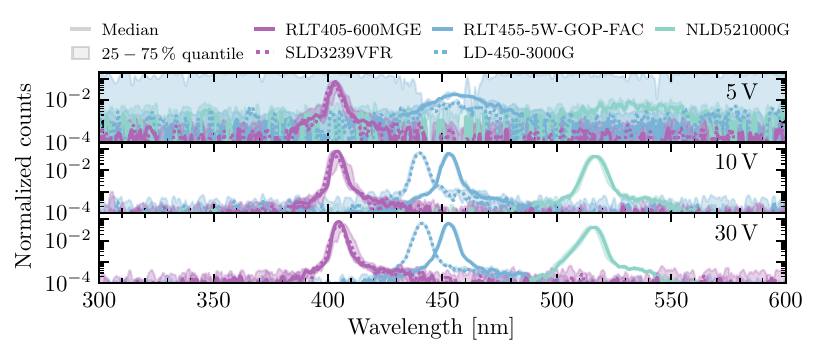}
    \end{subfigure}
    \caption{Photon yield (top), time profile (center), and spectrum (bottom) results for all P-CAL flasher channels that passed QC. Lines and shaded bands show the median and \unit[$25-75$]{\%} quantiles, respectively. The same photon yield fit of \cref{fig:flasher-summary} is used. For the time profile and spectrum, the continuous lines are obtained by linear interpolation of raw timing and spectral data, binned in \unit[20]{ps} and \unit[2]{nm} intervals, respectively. Top- and center-row figures are partly reproduced from~\cite{ghuman_situ_2025}.}
    \label{fig:pcal-flasher-summary}
\end{figure}

Similarly to the directional flashers, all 40 light pulser channels on the produced and assembled CBs were characterized and quality-checked with the measurement setup and parameters described in \cref{sec:directional}. Measurements on each channel were carried out over a bias voltage range of \unit[$0-30$]{V} and frequencies between \unit[100]{Hz} and \unit[100]{kHz}. All channels passed QC above the main operational bias voltage of \unit[10]{V}. Results for the full population of production CB channels are given in \cref{fig:pcal-flasher-summary}.
\par
\begin{figure}[h!]
    \centering
    \vspace{-12pt}
    \includegraphics[width=.99\textwidth]{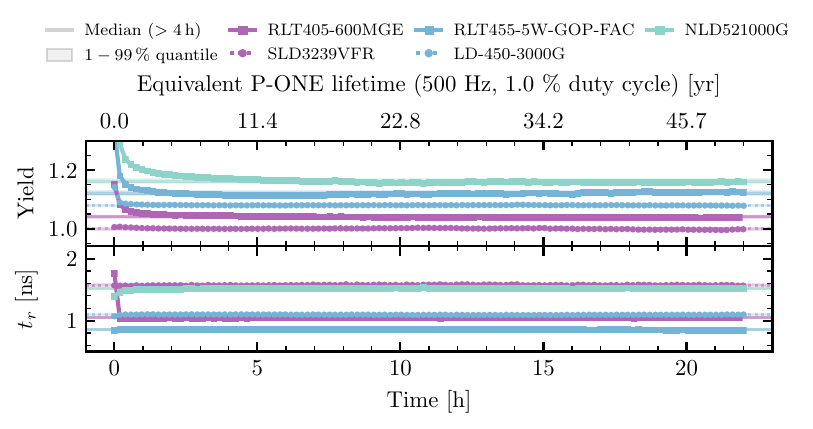}
    \vspace{-6pt}
    \caption{Aging test results for one of each CB flasher channels. Shown are the relative light yield (offset for diode types) and rise time of the observing SiPM for flasher pulses as a function of total operating time. Flashers were operated at a bias voltage of \unit[24]{V} and a frequency of \unit[100]{kHz}. Yield medians of different channels have been artificially offset for better visual separation.}
    \label{fig:pcal-aging}
\end{figure}

%\noindent
P-CAL flashers are also designed for a target lifetime of \unit[20]{years}. Therefore, one CB which passed QC was subjected to the same longevity test as the directional flashers (\cref{fig:flasher-aging}). To simulate accelerated aging, each flasher channel on the CB is continuously operated for \unit[24]{hours}, and pulse properties are monitored using external sensors. The results in \cref{fig:pcal-aging} show a warm-up phase during which the yield decreases by about \unit[10]{\%} in the first \unit[$1-3$]{hours} (depending on GaNFET and diode type), but no long-term degradation. All QC  measurements were taken without warm-up time. 
\par
\begin{figure}[h!]
    \centering
    \vspace{-6pt}
    \includegraphics[width=.99\textwidth]{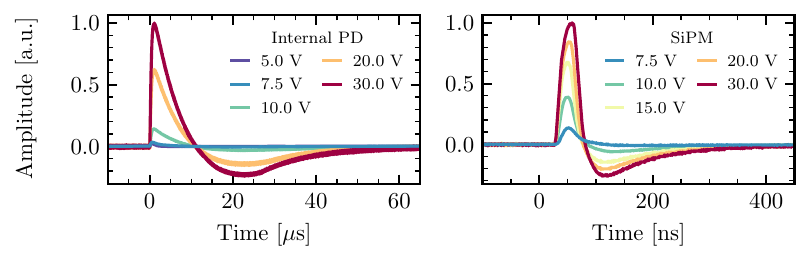}
    \caption{Example pulses of the internal photodiode (left) and SiPM (right) charge-monitoring channels taken with an oscilloscope. Colored curves indicate incident light pulses by the default \unit[405]{nm} diode on the CB at various bias voltages.}
    \label{fig:pcal-sensor-pulses}
\end{figure}

\begin{figure}[h!]
    \centering
    \vspace{-6pt}
    \includegraphics[width=.9\textwidth]{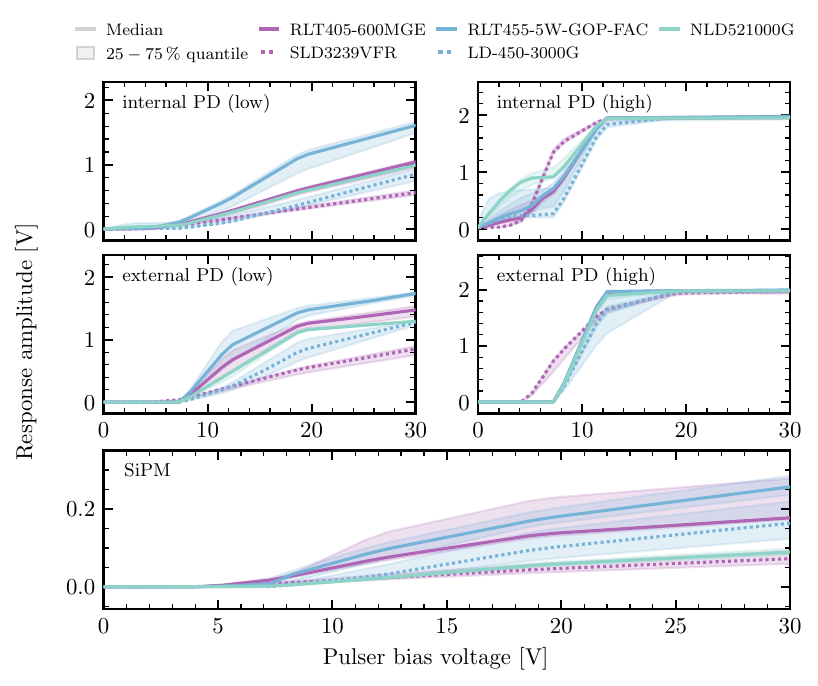}
    \caption{Results for all CB self-monitoring channels that passed QC. Shown are the sensor amplitudes for all channels and emitter types as a function of the pulser channel bias voltage.}
    \label{fig:pcal-sensor-summary}
\end{figure}

\noindent
Using the same setup, the analog outputs of the self-monitoring sensors were tested. For this QC step, bias voltages for each flasher were varied, and the analog outputs of the self-monitoring sensors were recorded with an oscilloscope. Notably, the absolute light scale will differ slightly when fully integrated into the P-CAL, but the relative sensor response to varying light yields can be characterized. Example output pulses for varying flasher bias voltages are shown in \cref{fig:pcal-sensor-pulses}. The measured amplitudes of the various sensor channels are used to estimate the incident optical power. Since these are used for in situ monitoring, we are interested in characterizing their response. By iteratively measuring sensor channel output amplitudes across different flasher channels and bias voltages, their relative behavior can be characterized. Results for all CBs are shown in \cref{fig:pcal-sensor-summary}.
%\par%\noindent
\par
\begin{figure}[h!]
    \centering
    \begin{subfigure}{.99\textwidth}
        \centering
        \vspace{-6pt}
        \begin{tikzpicture}
            \node[anchor=south west, inner sep=0] (img) at (0,0)
                {\includegraphics[trim=3.5cm 3cm 3.75cm 3cm, clip, width=.95\textwidth]{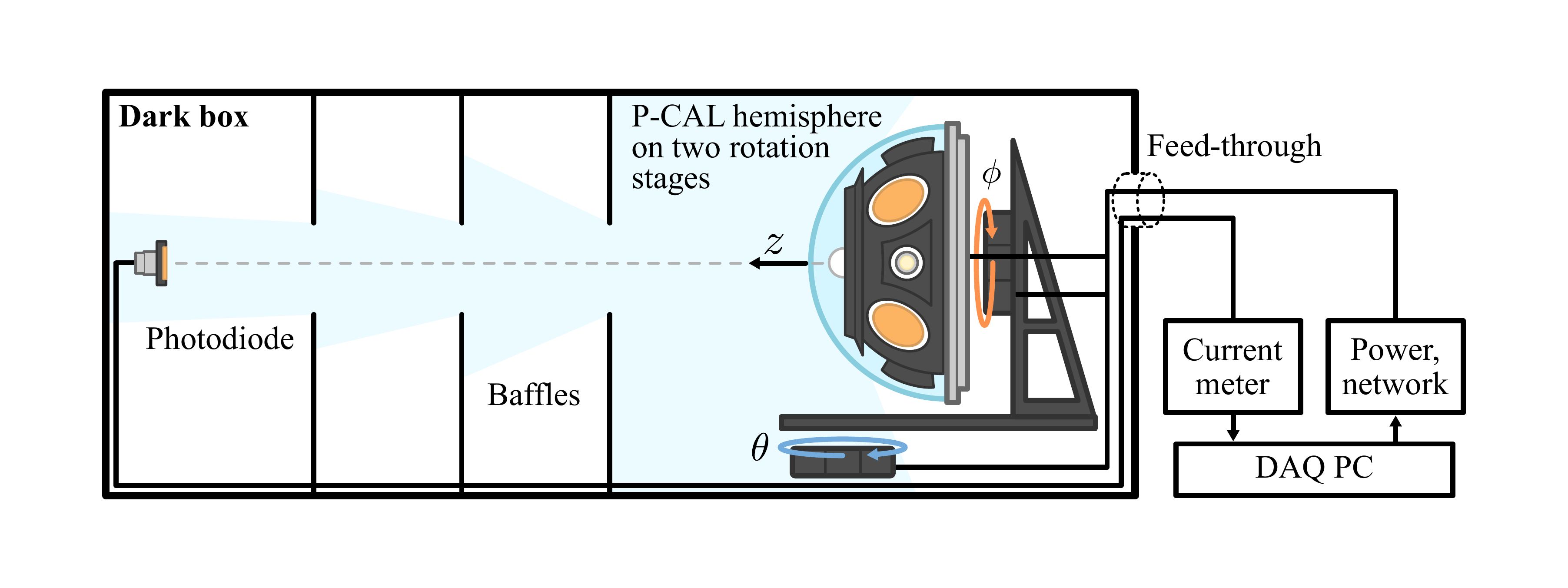}};
            \begin{scope}[x={(img.south east)}, y={(img.north west)}]
                \node at (0.875,0.99) {(side view)};
            \end{scope}
        \end{tikzpicture}
        \caption{Integrated P-CAL measurement setup.}
        \label{fig:pcal-setup}
    \end{subfigure}%
    \vspace{2mm}
    \begin{subfigure}{.99\textwidth}
        \centering
        \includegraphics[width=0.99\textwidth]{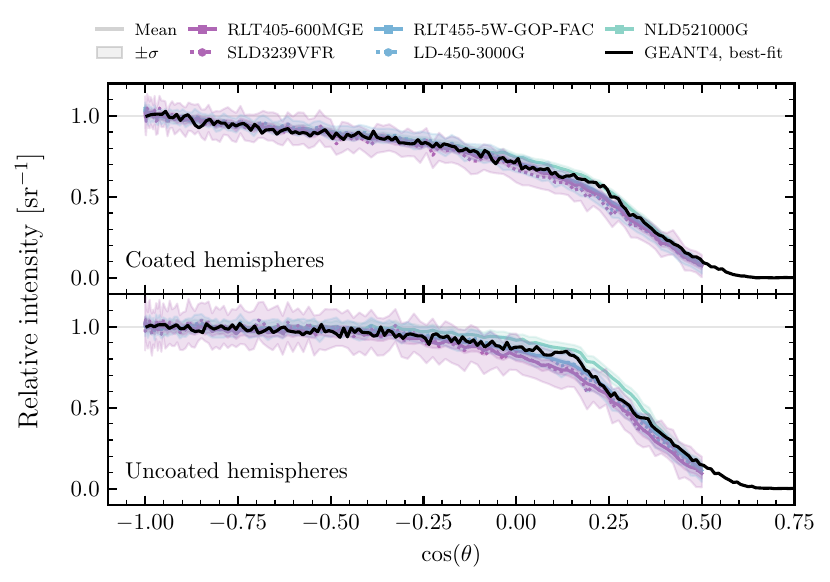}
        \caption{Measured P-CAL emission profiles in air.}
        \label{fig:pcal-profile}
    \end{subfigure}
    \caption{\textbf{(a)} Measurement setup for in-air P-CAL emission profile measurements using two rotation stages for scanning zenith and azimuth solid angles and a photodiode. \textbf{(b)} P-CAL emission profiles for two uncoated and two coated hemispheres measured in air using the two-axis rotation setup. Shown are the zenith profiles corresponding to each CB flasher channel averaged between hemispheres, and the best-fit GEANT4 simulation.}
    \label{fig:pcal-setup-and-profile}
\end{figure}

%\noindent
In-air characterization of integrated P-CAL emission profiles and self-monitoring sensors is done in a dedicated setup. This comprises a \unit[2.5]{m} long dark box with two heavy-duty rotation stages (Zaber X-RST120AK-E03) on one side, and a photodiode on the other. The stages allow two-axis rotation around the diffuser and can rotate a fully integrated P-CAL hemisphere. Taking incremental photodiode readings enables sampling zenith and azimuth solid angles of \unit[$0 - 130$]{degrees} and \unit[$0 - 360$]{degrees}, respectively.  The experimental setup is visualized in \cref{fig:pcal-setup}. At each point, flasher channels were enabled sequentially, and photodiode readings were taken using the picoammeter. The dark-corrected photocurrent is proportional to the incident light power. 

Four fully-integrated P-CAL hemispheres were produced, of which two are equipped with ClearSignal$^\mathrm{TM}$ anti-biofouling coating on the outer glass surface to reduce sub-sea organism growth~\cite{aghaei_long_2025}. Results for the measured emission profiles are shown in \cref{fig:pcal-profile} together with the corresponding best-fit GEANT4 simulation.
\par\noindent

\subsection{Integration}
As for the P-OM, all P-CAL components are integrated into the hemisphere using an adjusted version of the internal P-OM mounting frame, with mechanical interfaces for the diffusers, external self-monitoring sensors, and the camera system. 

\begin{figure}[h!]
    \centering
    \includegraphics[width=.75\textwidth]{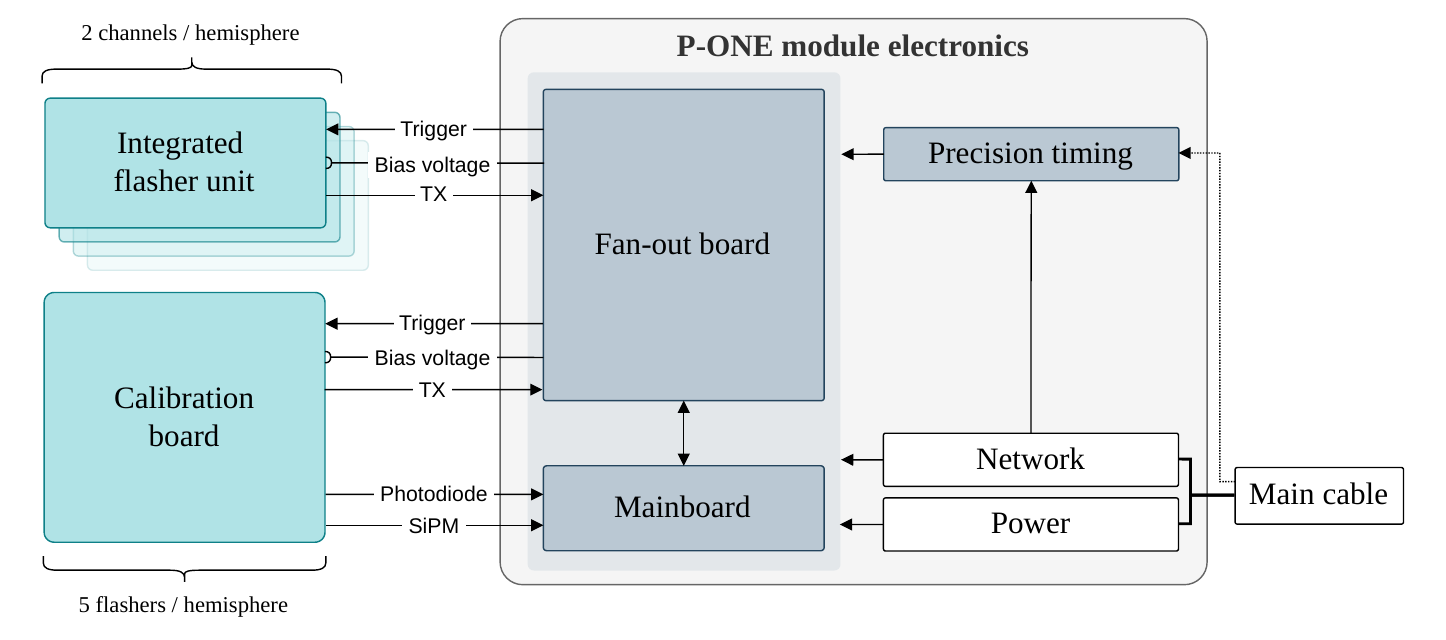}
    \caption{Schematic flowchart of the P-CAL integration into P-ONE module electronics.}
    \label{fig:flow-pcal}
\end{figure}

Electronically, the P-CAL hemisphere is identical in layout and components to a regular P-OM hemisphere, as discussed in \cref{sec:directional}. Thus, each hemisphere hosts a fan-out board, and one hemisphere in each module hosts the mainboard. The distinction is the replacement of four PMT signal connections with the self-monitoring sensors and the camera connection via USB. As for the P-OM, all analog channels are digitized on the mainboard using its ADC and TDC. The fan-out boards in each hemisphere both connect to the mainboard and provide all necessary connections to all components in each hemisphere. Supply and bias voltages for the CB are provided from the fan-out boards. Since only eight flasher channels are available per hemisphere, of which the CB takes up five, each P-CAL hemisphere hosts one upward- and one downward-facing directional flasher. Also, we use alternating special wavelengths to achieve redundancy with neighboring modules. The electronic integration flowchart for the P-CAL is sketched in \cref{fig:flow-pcal}. 
\par
\begin{figure}[h!]
    \centering
    \begin{subfigure}{.32\textwidth}
      \centering
      \includegraphics[width=.99\textwidth]{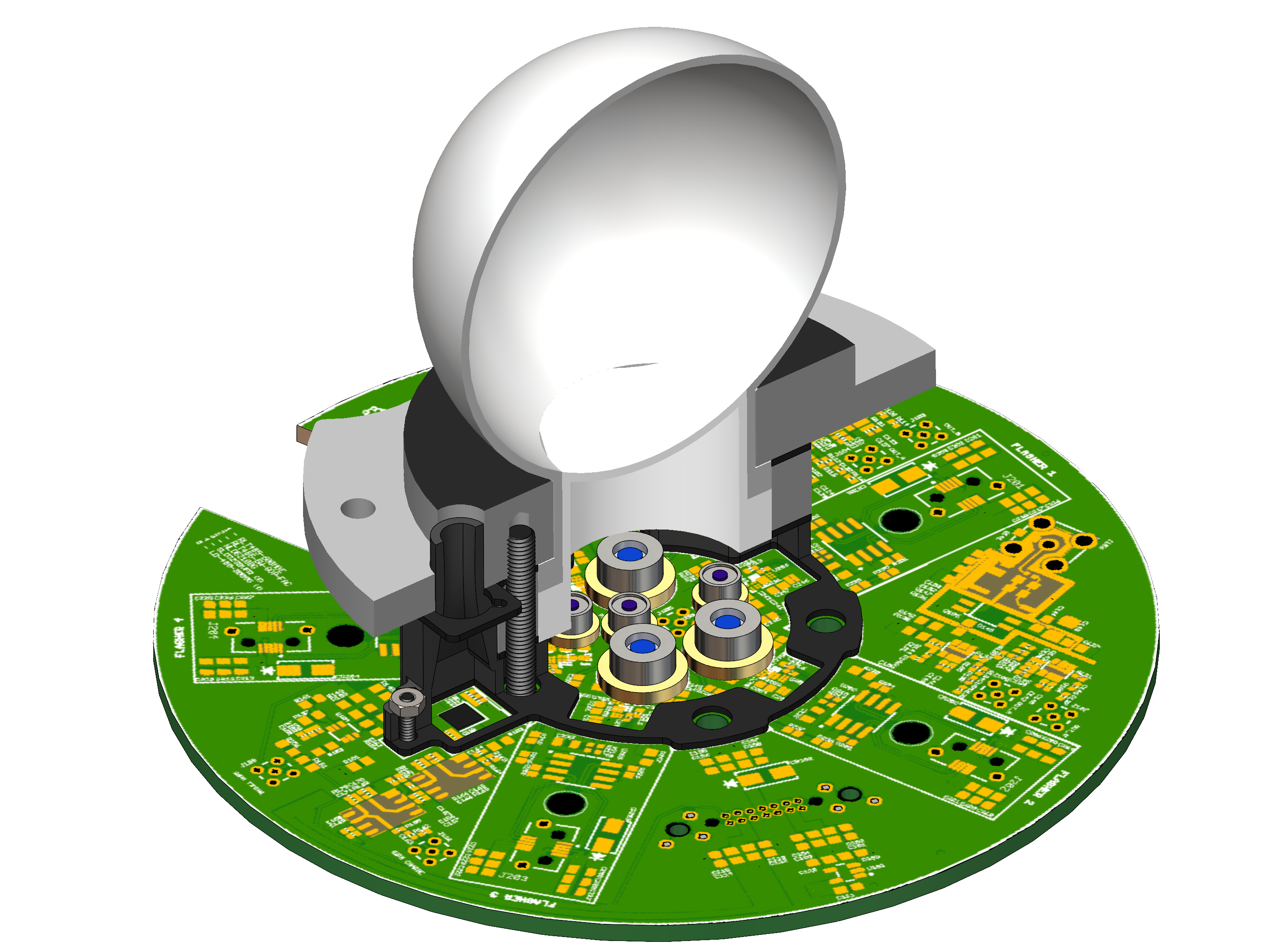}
      \caption{Diffuser stack}
      \label{fig:mechanical-pcal-a}
    \end{subfigure}%
    \hspace{1.5mm}
    \begin{subfigure}{.32\textwidth}
      \centering
      \includegraphics[width=.99\textwidth]{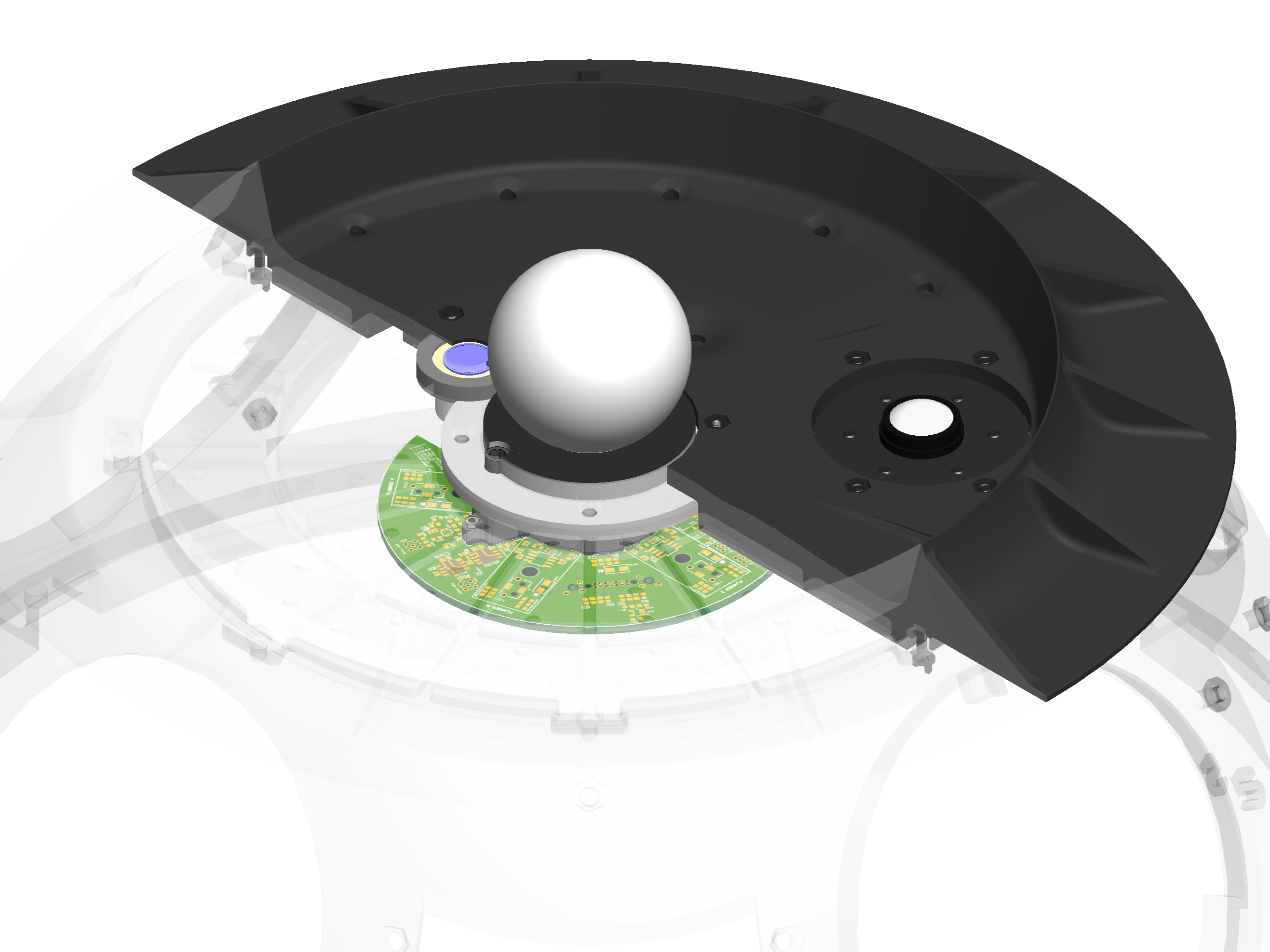}
      \caption{Upper volume}
      \label{fig:mechanical-pcal-b}
    \end{subfigure}
    \hspace{1.5mm}
    \begin{subfigure}{.32\textwidth}
      \centering
      \includegraphics[width=.99\textwidth]{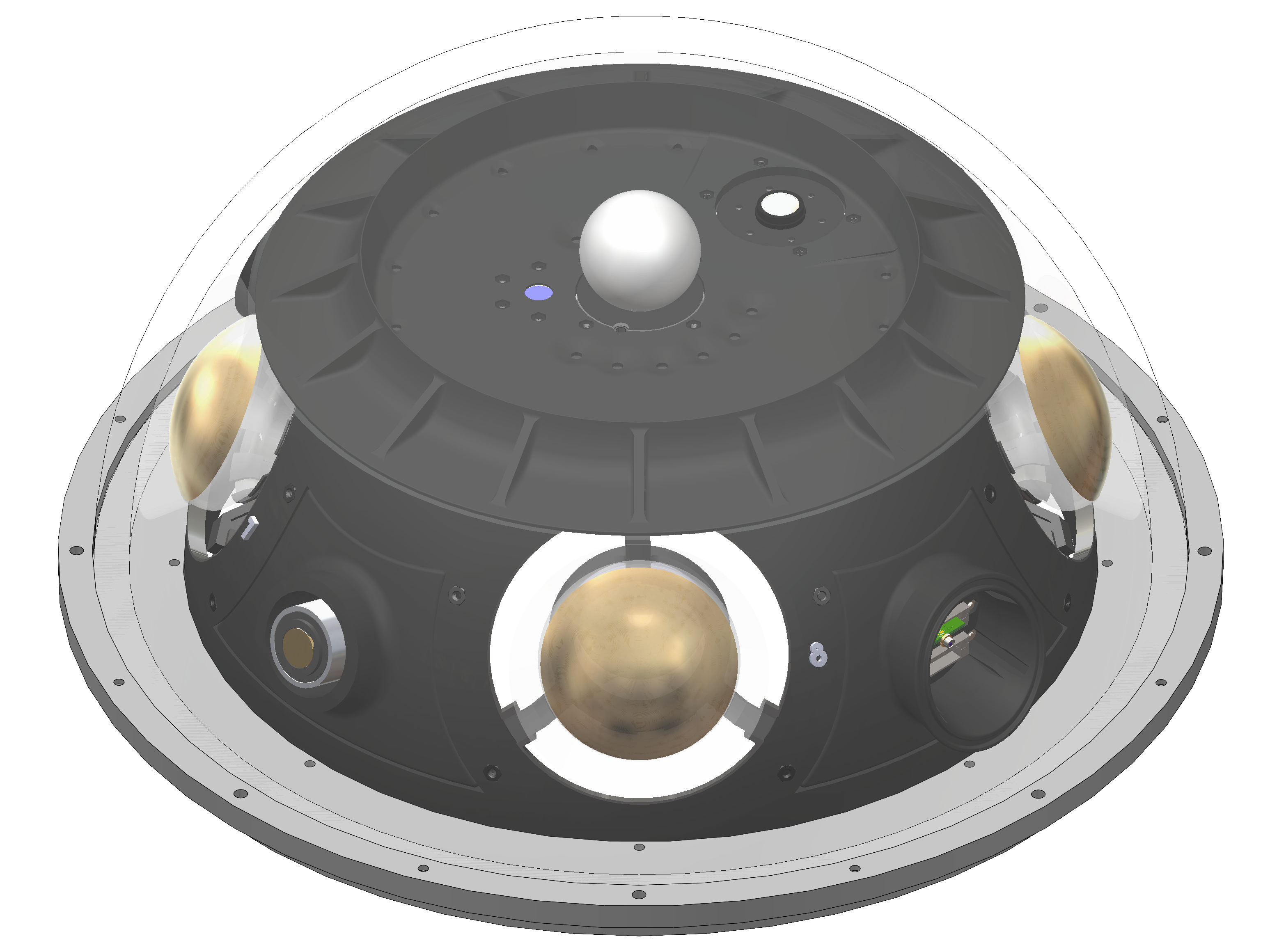}
      \caption{Integrated P-CAL}
      \label{fig:mechanical-pcal-c}
    \end{subfigure}
    % \hspace{1.5mm}
    % \begin{subfigure}{.23\textwidth}
    %   \centering
    %   \includegraphics[width=.99\textwidth]{graphics/pcal-production.png}
    %   \caption{P-CAL prototype}
    %   \label{fig:mechanical-pcal-d}
    % \end{subfigure}
    \caption{P-CAL \textbf{(a)} diffuser stack and \textbf{(b)} upper volume with shroud in section view, as well as \textbf{(c)} the fully-integrated P-CAL hemisphere (taken from~\cite{ghuman_situ_2025})}% and \textbf{(d)} the first prototype.}
    \label{fig:mechanical-pcal}
\end{figure}

\noindent
Mechanically, the assembly is embedded in optical gel, with approximately \unit[2.8]{liters} prepared in total, of which \unit[2.5]{liters} fill the hemisphere and are degassed prior to insertion of the mounting frame with the diffuser stack. Additional gel is added through the camera opening in the shroud to achieve the final fill level. The camera system is mounted through the shroud and secured to the frame, and the complete assembly is allowed to cure for \unit[24]{hours}. Following curing, the PMTs and fan-out board are installed, yielding a fully integrated hemisphere ready for line deployment (see \cref{fig:photos}). The integrated components and assembly stages are shown in \cref{fig:mechanical-pcal}.
\par\noindent
%

%%%% Simulation %%%%%%%%%%%
\section{Characterization in water}
\label{sec:water tank}
\begin{figure}[h!]
    \centering
    \begin{subfigure}{.49\textwidth}
      \centering
      \includegraphics[width=.85\textwidth]{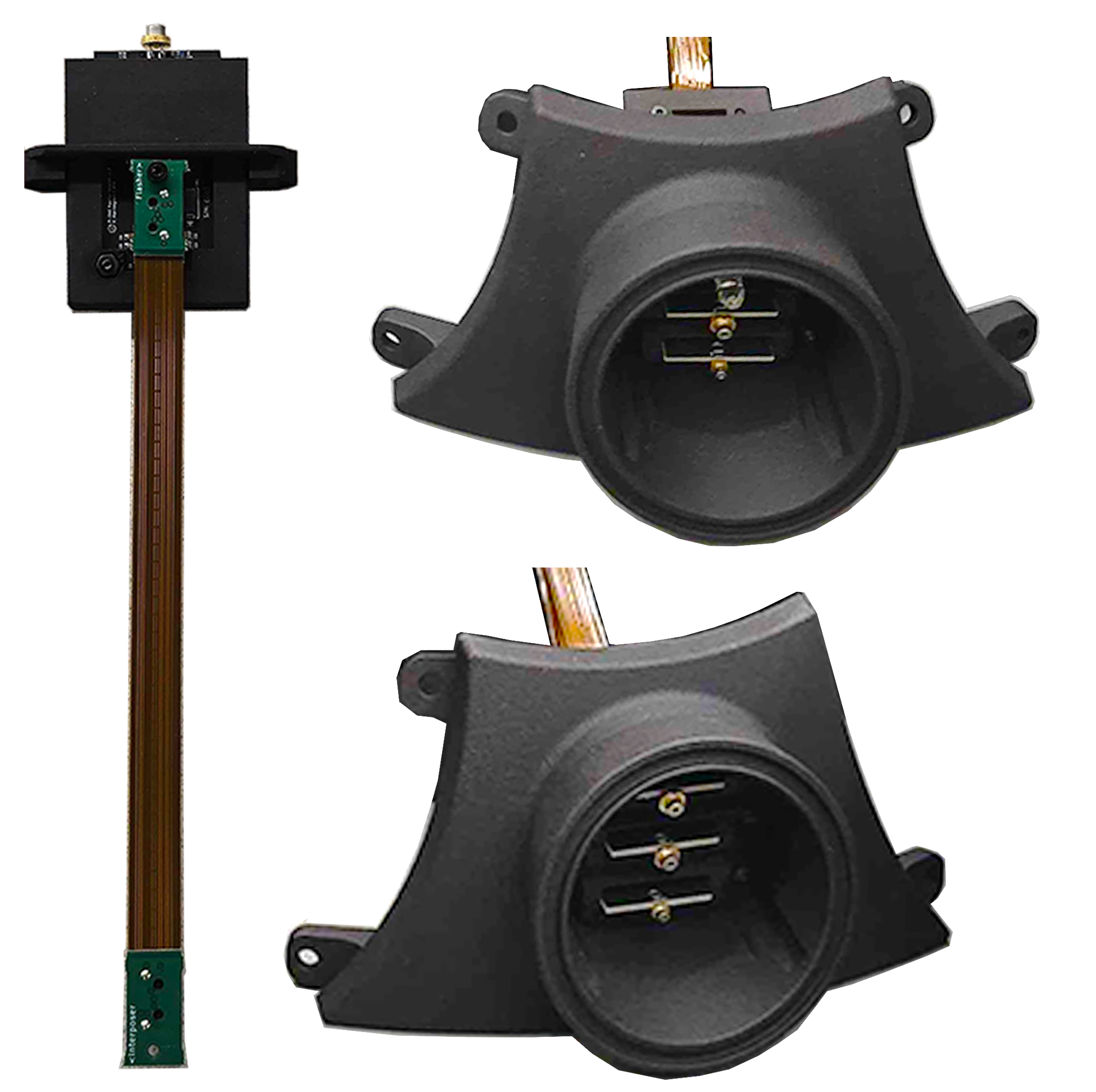}
      \caption{Directional flasher unit and assemblies}
      \label{fig:photos-a}
    \end{subfigure}%
    \hspace{1mm}
    \begin{subfigure}{.49\textwidth}
      \centering
      \includegraphics[width=.99\textwidth]{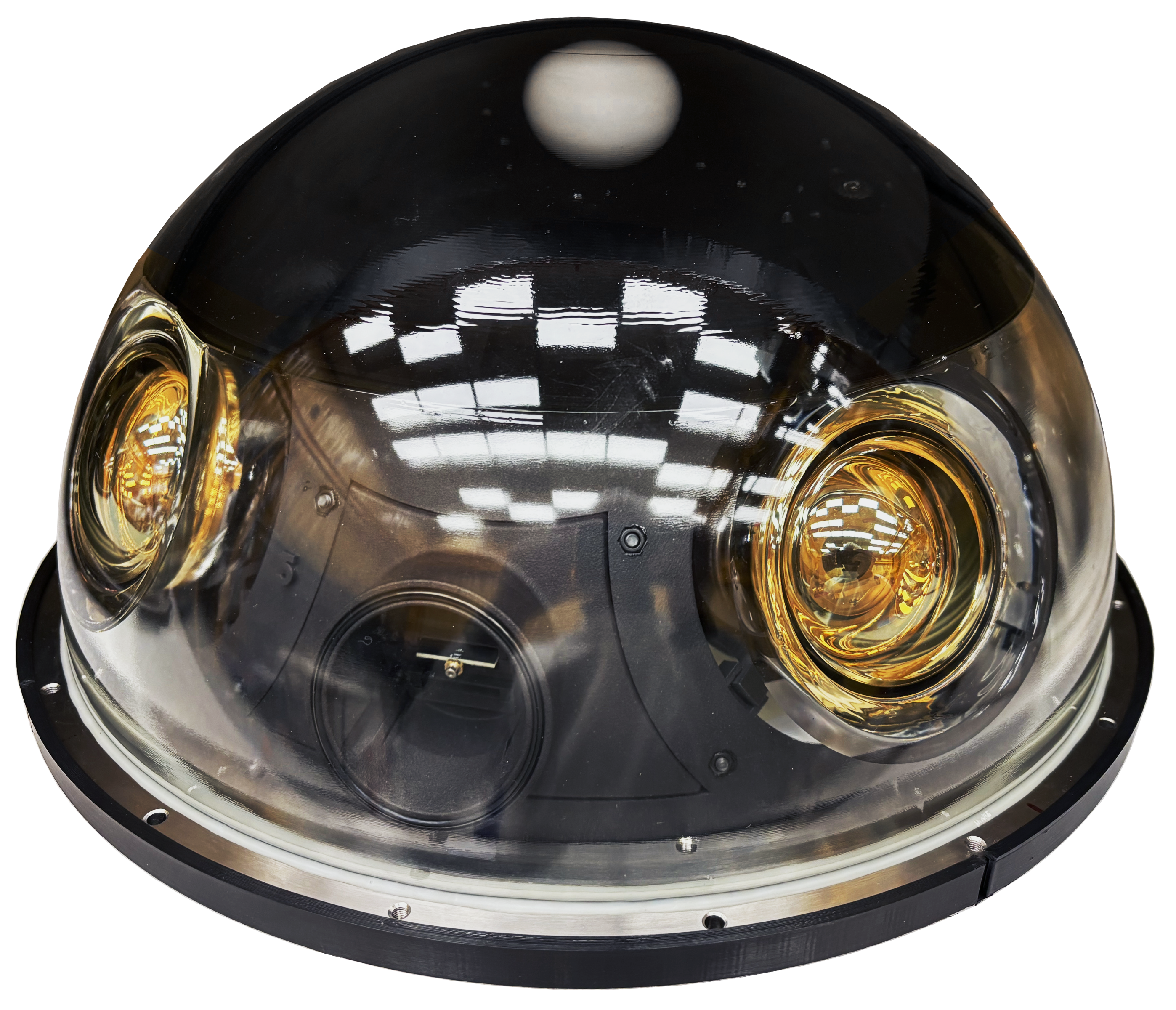}
      \caption{Complete P-CAL production hemisphere}
      \label{fig:photos-b}
    \end{subfigure}
    \caption{Photos of the complete optical calibration systems. \textbf{(a)} Directional flasher system showing an individual flasher (left), as well as LED-version (top right) and laser-version (bottom right) flasher unit assemblies; \textbf{(b)} a P-CAL production hemisphere ready for integration into the first P-ONE line. The prototype hemisphere used for underwater characterization is structurally identical to the production version.}
    \label{fig:photos}
\end{figure}

An integrated P-CAL hemisphere prototype (see \cref{fig:photos-b}) was tested in the Photosensor Test Facility (PTF), a water calibration tank operated at TRIUMF~\cite{gousy-leblanc_improving_2023}. This cylindrical tank measures approximately \unit[1.5]{m} in both diameter and height, and is equipped with a water-proof, movable gantry system and a water filtration system.

\subsection{Experimental setup}
\begin{figure}[h!]
    \centering
    \begin{subfigure}{.48\textwidth}
        \centering
        \begin{tikzpicture}
            \node[anchor=south west, inner sep=0] (img) at (0,0)
                {\includegraphics[width=.99\textwidth]{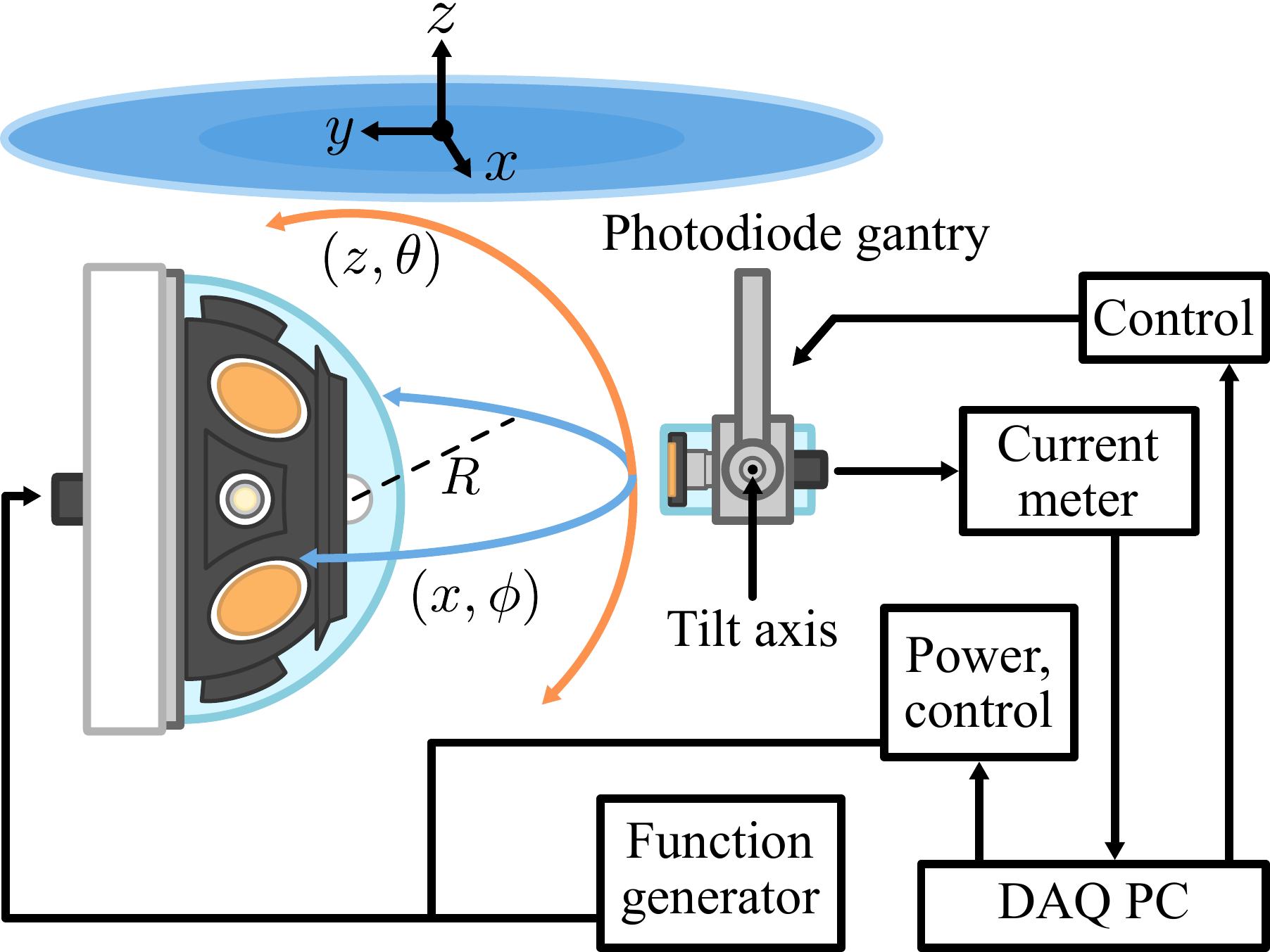}};
            \begin{scope}[x={(img.south east)}, y={(img.north west)}]
                \node at (0.875,0.85) {\footnotesize (side view)};
            \end{scope}
        \end{tikzpicture}
        \caption{P-CAL measurement setup.}
        \label{fig:ptf-tank-a}
    \end{subfigure}
    \hspace{1.5mm}
    \begin{subfigure}{.48\textwidth}
        \centering
        \includegraphics[width=.99\textwidth]{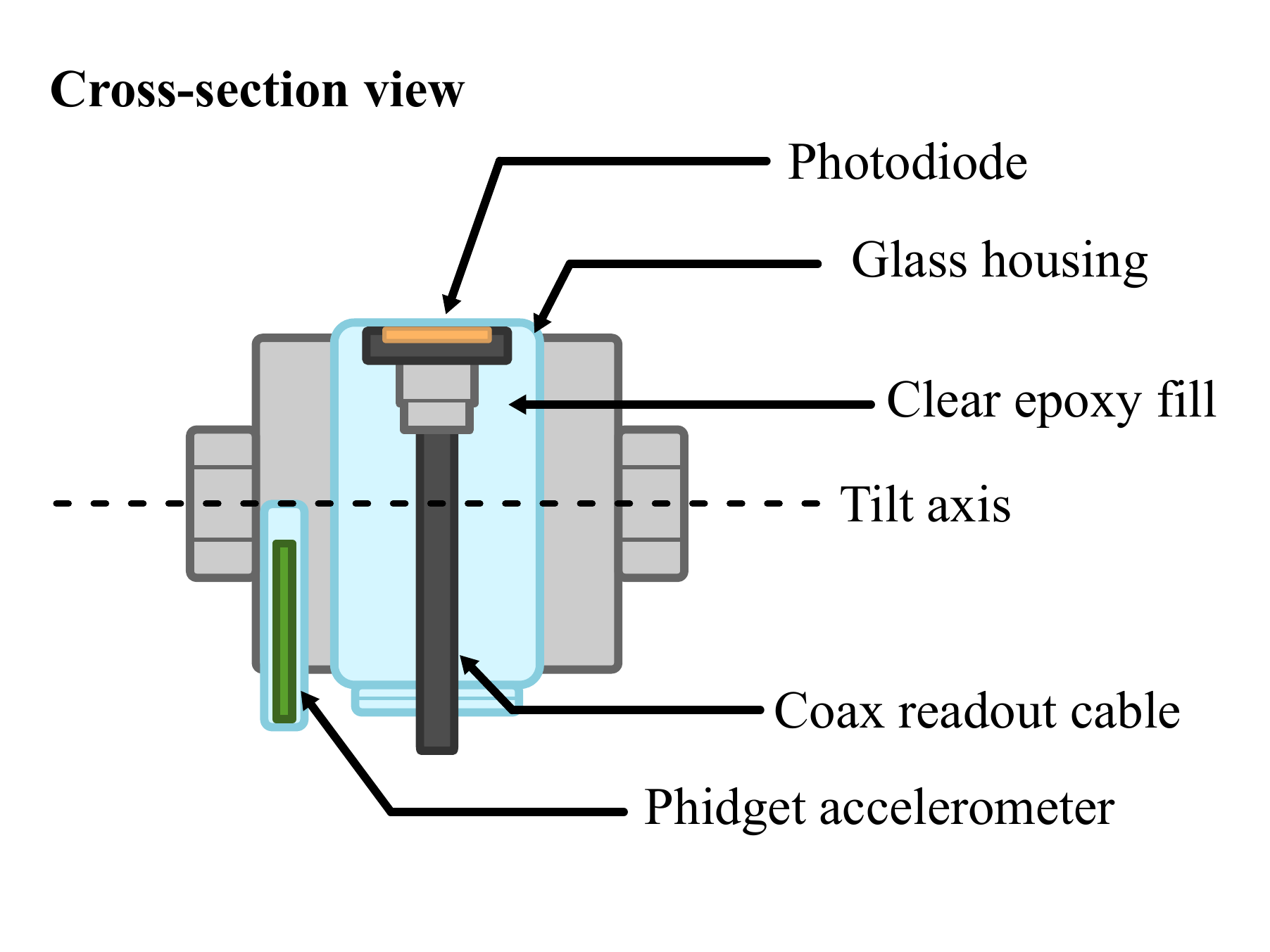}
        \caption{Internal layout of the gantry head.}
        \label{fig:ptf-tank-b}
    \end{subfigure}
    \caption{Water tank measurement setup for the  P-CAL hemisphere using  \textbf{(a)} the water tank of the PTF at TRIUMF and \textbf{(b)} the layout of the optical box on the gantry head. The movable gantry head with an integrated photodiode is used to measure light intensities at various radii $R$ around a predefined center. A DAQ computer controls the measurement system and the P-CAL.}
    \label{fig:water-setup}
\end{figure}

Originally designed for photosensor characterization, the tank features a gantry system capable of three-dimensional motion ($x$, $y$, $z$) and limited, two-axis head rotation ($\theta$, $\phi$). The gantry head carries an aluminum frame holding a glass housing filled with clear epoxy. Light detection is performed with a photodiode (Hamamatsu S2281-01) that is submerged in cured, transparent epoxy. It operates in photovoltaic mode, with a coaxial cable connection to an external picoammeter (Keithley 2502) for current monitoring. The aluminum frame also features a Phidget MOT1102\_0 accelerometer to measure the gantry head's tilt. With this, we can characterize the P-CAL emission profile as shown in \cref{fig:water-setup}. The gantry is moved along radial trajectories in the $(x,z)$ or $(x,y)$ planes, centered on the diffuser within the hemisphere. At each position, the photodiode records the detected photocurrent. The primary challenge is accurate alignment. This is addressed by performing linear sweeps along the gantry’s $y$ and $z$ axes while independently rotating the gantry head at each step. Cosine functions are fit to the resulting stepwise profiles (see \cref{fig:ptf-calibration}) to identify the global maximum in each scan, yielding the gantry head position and orientation that maximizes the photocurrent. This configuration corresponds to radial alignment toward the diffuser in the instrument’s forward direction.
\par\noindent
We use a prototype P-CAL hemisphere, with the mainboard replaced by a Raspberry Pi for simplicity. The latter interfaces our system to the outside via Ethernet and provides full control of all fan-out board functionality. Using the same components as in the P-ONE acoustic system ocean test~\cite{agostini_prototype_2025}, the instrument is sealed and waterproof by mounting it on a Delrin backplate. This backplate resembles the landing footprint of the P-ONE detection line cable and hosts a deep-sea cable connector providing a waterproof electronic interface. The matching cable measures \unit[10]{m} and contains ten copper wires. Eight of these comprise four twisted-pair lines, and two are higher-gauge power wires. Two pairs provide an Ethernet connection to the Raspberry Pi, the two power wires carry a supply voltage of \unit[30]{V} to the internal DC/DC converter, and one of the remaining pairs is used for providing light pulse trigger signals into the hemisphere (refer to \cref{fig:flow-pcal}). Mechanically, the P-CAL is mounted in the tank with its z-axis oriented horizontally (see \cref{fig:water-setup}). This is achieved with a dedicated L-bracket assembly fastened to the Delrin backplate and fixed to the base of the tank. The diameter of the water tank, the gantry collision detection, and the P-CAL assembly structure allow for measurement radii of up to approximately \unit[35]{cm} from the center of the diffuser. The gantry head further includes tilt sensors that allow correcting for deviations from its programmed lateral tilt angle, $\theta$, during the scan. All controls and data storage are handled through a custom MIDAS software framework~\cite{triumf_maximum_2015}, and data are stored on disk for offline processing.

A full measurement scan comprises a geometrically defined sweep path stepping radially around a defined coordinate center in either the $(x,y)$ or $(y,z)$ plane, recording $\mathcal{O}(25)$ photodiode current readings per step. The corrected photo current is obtained by subtracting the sensor's dark-current offset, which is measured intermittently with the light source disabled. Light pulses are triggered using a function generator (Tektronix AFG31152) with repetition frequencies between \unit[$10$]{kHz} and \unit[$30$]{kHz}. The photodiode current is integrated over a few milliseconds, and each current reading thus represents the total optical power deposited on the sensor by multiple light pulses. 

\subsection{Results}
\paragraph*{Alignment}
Before performing emission profile measurements, the P-CAL position in the gantry's coordinate system and the gantry head rotation need to be calibrated. As a first step, the alignment laser included with the gantry head was used to manually position the P-CAL mounting structure parallel to the gantry coordinate system's y-axis. In addition, carefully inching the gantry head towards the P-CAL in a small-step grid enabled collision detection to determine the glass hemisphere's position in space. Together, this provided a first estimate of the pressure housing's physical position in gantry coordinates. Given the known instrument dimensions, these can be translated into diffuser coordinates, specifying the center point for radial scans. In a second step, the initial coordinates are used to perform a more precise alignment. Here, the P-CAL is continuously flashing, and linear steps and rotations of the gantry in both the $(x, \phi)$ and $(z, \theta)$ parameters are performed. With photodiode readings taken at each step, this calibrates the linear axis position of the diffuser and the radial gantry head rotation in both angles ($\theta$, $\phi$).

\begin{figure}[h!]
    \centering
    \vspace{-6pt}
    \includegraphics[width=.99\textwidth]{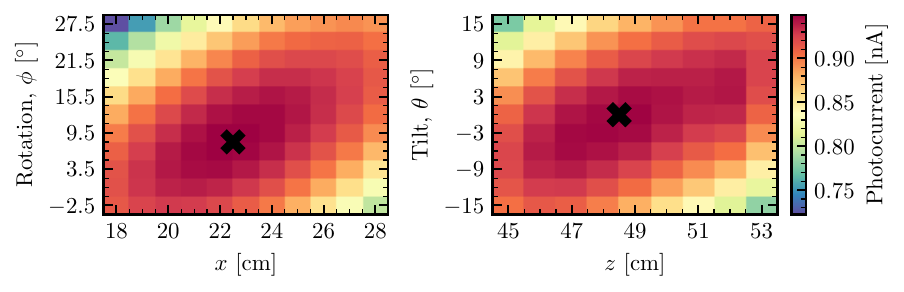}
    \caption{Water alignment measurements for the gantry system with a flashing P-CAL in $(x,\phi)$ and $(z,\theta)$. Both figures show the measured photocurrent (color) as a function of the linear step coordinate and the respective gantry head rotation angle. Optimal parameters are marked.}
    \label{fig:ptf-calibration}
    \vspace{-6pt}
\end{figure}

Results of this sequence for the mentioned gantry alignment are shown in \cref{fig:ptf-calibration} for measurements in water. The figure shows axes and rotation-alignment scans in both the $(x,\phi)$ and $(z,\theta)$ coordinates, and the optimal parameters found are used as calibrated reference values for all subsequent emission profile measurements. We estimate this to achieve angular and positional errors of less than \unit[5]{degrees} and \unit[5]{mm} in each axis, respectively.
\par\noindent

\paragraph*{Emission profile scan}
After alignment calibration, emission profile scans were performed with the gantry system in both air and water. Performing the scan in both serves to verify the independent in-air measurements using the \unit[2.5]{m} rotation setup (see \cref{sec:pcal:characterization}) and to match experimental data across different configurations using the GEANT4 simulation framework.

\begin{figure}[h!]
    \centering
    \includegraphics[width=.95\textwidth]{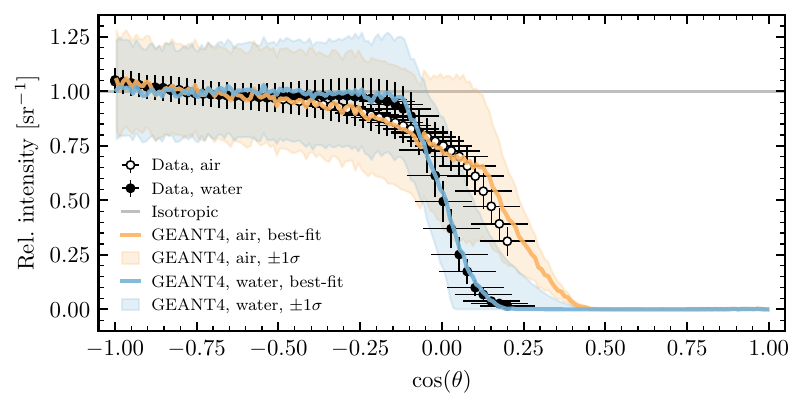}
    \caption{Emission profile data of the P-CAL hemisphere in air and water using a scanning radius of \unit[$R=0.33$]{m}. Data have been averaged across horizontal $(x, \phi)$ and vertical $(z, \theta)$ sweep paths and diode channels, and errors indicate the resulting standard deviation. Also shown is the interpolated best-fit GEANT4 simulation and its variation due to close proximity and tolerances.}
    \label{fig:ptf-air-water}
\end{figure}

For water measurements, the tank is filled with about \unit[300]{gallons} (\unit[$\sim$1136]{liters}) of deionized water; for air measurements, it is completely drained. In both cases, a sweep radius of about \unit[0.33]{m} is used for the gantry path, and profiles on the horizontal and vertical planes are recorded. Both scenarios are simulated in GEANT4 and corrected for the gantry head's limited angular acceptance by selecting recorded photons based on their incident angles. Measurement results and GEANT4 simulations are shown in \cref{fig:ptf-air-water}, normalized to the average of $\cos(\theta) \in [-1, -0.5]$. 
\par%\noindent
In general, we find remarkable agreement between data and simulation, especially given the geometric tolerances of the system. However, the observed spread in the simulation datasets, given the expected parameter variations (shaded bands), is significant. These arise from near-field effects caused by the sensor’s proximity to the light source and from geometric tolerances in both the instrument and the measurement system. The resulting profiles indicate that the optical gel functions as expected and the matching refractive index in water provides an optimized transition region near $\cos(\theta)=0$. This is because glass and gel become largely transparent in water, and the diffuser acts as the light source. In air, the refractive-index difference between air and gel causes strong lensing of the diffuse light, especially in the transition region, making the entire hemisphere act as an effective light source.

\begin{figure}[h!]
    \centering
    \includegraphics[width=.99\textwidth]{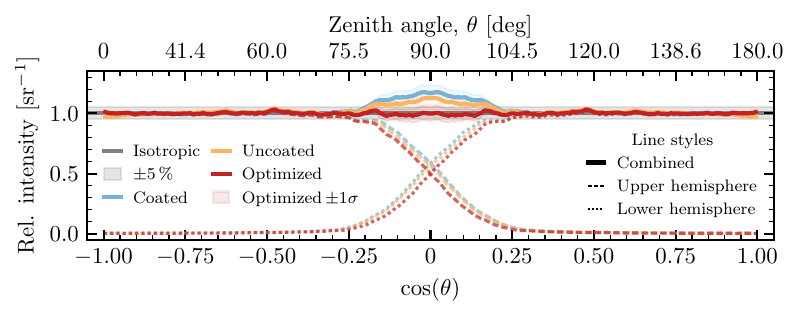}
    \caption{P-CAL emission profiles at \unit[50]{m} distance in water simulated with GEANT4. The profiles shown include surface parameterizations that match in-air measurements for both optimized simulation parameters (see \cref{fig:pcal-geant4}) and coated and uncoated hemispheres (see \cref{fig:pcal-profile}). Error bands indicate simulated standard deviations resulting from a subset of simulated design tolerances. Water scattering was modeled as Mie-dominated with an asymmetry parameter of $g=0.92$ and a geometric scattering length of \unit[75]{m}; the absorption length was set to approximately \unit[30]{m}. Limited simulation statistics were smoothed using a running average with window size $N=3$.}
    \label{fig:pcal-final}
\end{figure}

This agreement between data and simulation indicates that our GEANT4 framework can be used to estimate the P-CAL far-field performance in seawater. Using these results, we simulated the emission profile of a deployed P-CAL module in P-ONE. For this, we increase the detection distance in GEANT4 to \unit[50]{m}, matching the smallest module distance on the first P-ONE line, and add a geometric scattering length of \unit[75]{m} to the optical water properties~\cite[e.g.~][]{smith_optical_1981}; the absorption length was set to approximately \unit[30]{m} to match expectations from STRAW measurements. We modeled scattering as Mie-dominated and employed the Henyey–Greenstein parameterization~\cite{henyey_diffuse_1941} in GEANT4, using an asymmetry parameter of $g=0.92$ for the expected forward-peaked scattering in ocean water~\cite[e.g.~][]{freda_revisiting_2012, balasi_method_2018}. \Cref{fig:pcal-final} shows simulated profiles normalized to the average of $\cos(\theta) \in [-1, -0.5]$ at \unit[50]{m} detection distance. These profiles comprise the best-case isotropic design (refer to \cref{fig:pcal-geant4}) as well as for the simulation parameters best-fitting to the measured profiles of coated and uncoated hemispheres (refer to \cref{fig:pcal-profile}). 
\par%\noindent
We find that all measured P-CAL design realizations exhibit an excellent isotropy grade in the forward direction (zenith angle ranges $[0^\circ, 75^\circ] \cup [105^\circ, 180^\circ]$) at $\unit[1.00 \pm 0.01]{}$ for all of the coated, uncoated, and best-case design parameterization of the first production run, respectively, and assuming symmetry between hemispheres. As expected, the transition region ($75^\circ < \text{zenith angle} < 105^\circ$) shows sensitivity to geometric tolerances, leading to small upward fluctuations in the combined emission profiles. The corresponding isotropy grades in this zenith range average to $1.11 \pm 0.05$, $1.07 \pm 0.03$, and $0.99 \pm 0.01$ for the same order of profile realizations as described above. Over the full angle range, we find respective isotropy grades of $1.03 \pm 0.05$, $1.02 \pm 0.04$, and $1.00 \pm 0.01$. These results demonstrate the full applicability of the developed P-CAL modules for the first P-ONE line; however, they also prompt possible optimizations of the transition-region design to achieve full-zenith isotropy by marginally tuning the internal instrument geometry. These will be implemented in second-generation instruments for future detection lines. 

The impact of transition-region anisotropies on calibration measurements is primarily governed by uncertainties in the relative positions of emitters and receivers. Such positional offsets translate into uncertainties in the sampled angular range of the P-CAL emission profile and are therefore most relevant in the transition regime, where the emission intensity varies with angle. A conservative estimate of this effect corresponds to a \unit[10]{\%} variation over an angular span of approximately \unit[15]{deg}. Assuming a linear dependence within this range, an angular mismatch of one degree introduces a systematic uncertainty of about \unit[0.7]{\%} in measurements making use of the intensity scale (e.g., water attenuation or PMT efficiencies). Since relative positioning between modules is expected to be stable well below the meter scale, the resulting contribution to the overall uncertainty is limited to a few percent. Timing-synchronization measurements are insensitive to these variations.

Long-term operation and reliability of the P-CAL are primarily determined by the extreme deep-sea environment. No degradation of the light-pulse driver electronics is expected over the anticipated lifetime of P-ONE (\cref{sec:pcal:characterization}), while the remaining integrated electronics (see \cref{fig:flow-pcal}) are part of the common data acquisition infrastructure designed for a target lifetime of \unit[20]{years}. The titanium–glass pressure housing ensures stable operation at water depths of \unit[$1600 - 2700$]{m}. The dominant factors influencing long-term performance are sedimentation and biofouling on the glass hemispheres. Both processes lead to a gradual, time-dependent reduction in emitted light intensity~\cite{aghaei_long_2025}. To study the mitigation of these effects, an anti-biofouling coating is applied to one of the P-CALs. The glass surface will be continuously monitored by the integrated camera system to assess the coating's effectiveness. Nevertheless, ongoing calibration will be required, and data from the first deployed line will guide the selection of effective mitigation strategies for future instruments. Consequently, the first year of data-taking is expected to be especially valuable for calibration, as the impact of sedimentation and biofouling should still be minimal then.

%%%% SUMMARY %%%%%%%%%%%
\section{Summary}
\label{sec:summary}
In this article, we describe the design and performance of the optical calibration systems for P-ONE. These comprise directional and isotropic light-pulsers, and we detail their electrical and optical designs, their functional and opto-mechanical integration into the P-ONE detector infrastructure, and measured performance characteristics using experimental setups. For the developed P-CAL instruments, we also conducted dedicated in-air and in-water measurements of the angular emission profile at the PTF facility located at TRIUMF.

\par%\noindent

For the directional flashers, we designed economical nanosecond light pulsers with LED and laser-diode emitter wavelengths between \unit[$365-520$]{nm}. A total of 350 flashers for integration into the first P-ONE line were produced and characterized, achieving pulse FWHMs as narrow as \unit[1.4]{ns} and light intensities up to \unit[$10^{10}$]{photons per pulse}. The flasher design is complemented by a dedicated mechanical system that integrates flashers into all optical modules in P-ONE. For the isotropic P-CAL instrument, the same directional pulser technology was adapted to deliver bright, self-monitoring light pulses within the same spectral range as the directional system. Here, eight boards with a total of 40 light-emission channels were produced and characterized, achieving pulse FWHMs as narrow as \unit[1.5]{ns} and light intensities up to \unit[$10^{11}$]{photons per pulse}. These boards also include self-monitoring readout channels for two photodiodes and a SiPM, providing in situ monitoring of all emitted light pulses. Simulations and measurements confirm that the intricate optical design of our instruments, based on an optical diffuser, 3D-printed structures, and optical gel, can ultimately achieve a P-CAL far-field isotropy of $1.00 \pm 0.01$ over the entire zenith range. In the transition region where both hemispheres contribute, and tolerances add up, all measured hemisphere profiles show slight anisotropic over-fluctuations on the order of \unit[$5-10$]{\%}. This will be improved with design modifications for the next instrument generation. These results were obtained from in-air and in-water measurements as well as a dedicated GEANT4 simulation framework. In this process, we also performed a refractive index and attenuation length measurement of the used optical gel (Wacker SilGel 612), showing an approximately constant refractive index of \unit[1.4]{} and an attenuation length of \unit[$0.2-0.5$]{m$^{-1}$} over the spectral range of interest.

\section*{Acknowledgments}
\noindent We thank Ocean Networks Canada for the very successful operation of the NEPTUNE observatory, as well as the support staff from our institutions without whom this experiment and P-ONE could not be operated efficiently. We acknoledge support from the Canadian Department of Fisheries and Oceans. We acknowledge the support of the Natural Sciences and Engineering Research Council of Canada (NSERC) and the Canadian Foundation for Innovation (CFI). This research was enabled in part by support provided by the BC and Prairies DRI and the Digital Research Alliance of Canada (alliancecan.ca). This research was undertaken thanks in part to funding from the Canada First Research Excellence Fund through the Arthur B. McDonald Canadian Astroparticle Physics Research Institute. P-ONE is supported by the Collaborative Research Centre 1258 (SFB1258) and the Cluster of Excellence ORIGINS, funded by the Deutsche Forschungsgemeinschaft (DFG) under Germany’s Excellence Strategy, and by the European Research Council (ERC) under the European Union’s Horizon Europe programme through an ERC Advanced Grant. We acknowledge support by the National Science Foundation. This work was supported by the Science and Technology Facilities Council, part of the UK Research and Innovation, and by the UCL Cosmoparticle Initiative. This work was supported by the Polish National Science Centre (NCN).

%%%%%%%%%%%%%%%%%%%%%%%%%%%%%%%%%%%%%%%%%%%%%%%%%%%%%%%%%%%%%%%%%%%%%%%%%%%%%%%%
% bibliography
\printbibliography

\end{document}